\documentclass[manuscript,screen,noacm]{acmart}

\AtBeginDocument{%
  \providecommand\BibTeX{{%
    \normalfont B\kern-0.5em{\scshape i\kern-0.25em b}\kern-0.8em\TeX}}}

\setcopyright{none}

\acmConference[ACM Transactions on Quantum Computing]{}{2024}{NY}

\usepackage{xcolor}
\usepackage{physics}
\usepackage[procnumbered,linesnumbered,ruled]{algorithm2e}
\usepackage{amsmath,amssymb,amsfonts}
\usepackage{algorithmic}
\usepackage{graphicx}
\usepackage{hyperref}
\usepackage{subcaption}

\newcommand{\eat}[1]{{}}
\newcommand{\magentaa}[1]{\textcolor{black}{#1}}

\newcommand{\bluee}[1]{\textcolor{black}{#1}}
\newcommand{\para}[1]{\smallskip \noindent\textbf{#1}}
\newcommand{\softpara}[1]{\smallskip \noindent \underline{#1}}

\newcommand{\calb}{\mathcal{B}}
\newcommand{\vx}{\textbf{x}}
\newcommand{\iso}{{\tt ISO}\xspace}

\newcommand{\vsp}{\vspace{0.01in}}
\def\blackbox{\hfill {\vrule height6pt width6pt depth0pt}}
\newcounter{theorem}
\newtheorem{thm}{Theorem}
\newtheorem{lem}{Lemma}
\newtheorem{ob}{Observation}
\newtheorem{corollary}{Corollary}
\newtheorem{conj}{Conjecture}

\newenvironment{theorem}{\vsp \begin{thm} \nopagebreak}{{\hfill$\Box$} \end{thm} \vsp}
\newenvironment{thm-prf}{\vsp \begin{thm} \nopagebreak}{\end{thm}}
\newenvironment{prf}{\noindent {\bf {\em Proof:}} \nopagebreak }{{\hfill$\blackbox$}}
\newenvironment{lem-prf}{\vsp \begin{lem} \nopagebreak}{\end{lem}}
\newenvironment{observation}{\noindent \begin{ob}}{{\hfill$\Box$}\end{ob}}

\newenvironment{cor-prf}{\vsp \begin{corollary} \nopagebreak}{\end{corollary}}
\newenvironment{conjecture}{\noindent \begin{conj}}{{\hfill$\Box$}\end{conj}}

\newcommand{\mpfont}{\scriptsize}

\ifx\noeditingmarks\undefined
    \newcommand{\MPworker}[2]{{\color{#1}\vrule\vrule}{\marginpar{\color{#1}\mpfont #2}}}
\else
    \newcommand{\MPworker}[2]{}
\fi

\ifx\noeditingmarks\undefined
    
\else
    
\fi

\ifx\noeditingmarks\undefined
    
\else
    
\fi

\begin{document}

\title{Optimizing Initial State of Detector Sensors in Quantum Sensor Networks}

\author{Caitao Zhan}
\email{cbzhan@cs.stonybrook.edu}
\orcid{0000-0001-7236-9737}
\author{Himanshu Gupta}
\email{hgupta@cs.stonybrook.edu}
\orcid{0000-0001-9131-1530}
\affiliation{%
  \institution{Department of Computer Science, Stony Brook University}
  \city{Stony Brook}
  \state{NY}
  \country{USA}
}

\author{Mark Hillery}
\affiliation{%
  \institution{Department of Physics and Astronomy, Hunter College of the City University of New York}
  \city{New York}
  \country{USA}}
\email{mhillery@hunter.cuny.edu}
\orcid{0000-0002-1296-075X}

\renewcommand{\shortauthors}{C. Zhan, H. Gupta, M. Hillery}



\begin{abstract}
In this paper, we consider a network of quantum sensors, where each sensor is a qubit detector that ``fires,'' i.e., its state changes when an event occurs close by. 
The change in state due to the firing of a detector is given by a unitary operator, which is the same for all sensors in the network. 
Such a network of detectors can be used to localize an event, using a protocol to determine the firing sensor, presumably the one closest to the event.
The determination of the firing sensor can be posed as a {\em Quantum State Discrimination} problem, which incurs a probability of error depending on the initial state and the measurement operators used. 

In this paper, we address the problem of determining the optimal initial global state of a network of detectors that incur a minimum probability of error in determining the firing sensor. For this problem, we derive necessary and sufficient conditions for the existence of an initial state that allows for perfect discrimination, i.e., zero probability of error. 
Using insights from this result, we derive a conjectured optimal solution for the initial state, provide a pathway to prove the conjecture, and validate the conjecture empirically using multiple search heuristics that seem to perform near-optimally. 
\end{abstract}

\maketitle

\section{Introduction}

Quantum sensors, being strongly sensitive to external disturbances, can measure various physical phenomena with extreme sensitivity. 
These quantum sensors interact with the environment 
and have the environment phenomenon or parameters encoded in their state, which can then be measured.
\magentaa{Thus, quantum sensors can facilitate several applications, including gravitational wave detection, astronomical observations, atomic clocks, biological probing, 
target detection, etc.~\cite{photonic_quantum_sensing}.
A study~\cite{quantum_radar} has shown the advantages of microwave quantum radar in the detection of a target placed in a noisy environment by exploiting quantum correlations between two modes, probe and idler.}
Estimation of a single continuous parameter by quantum sensors can be further enhanced by using a group of entangled sensors, improving the standard deviation of measurement by a factor of $1/\sqrt{N}$ for $N$ sensors~\cite{Giovannetti_2011}. 
Generally, a network of sensors can facilitate spatially distributed sensing; e.g., a fixed transmitter's signal observed from different locations facilitates localization via triangulation. 
Thus, as in the case of classical wireless sensor networks, it is natural to deploy a network of quantum sensors to detect/measure a spatial phenomenon, and there has been recent interest in developing protocols for such quantum sensor networks (QSNs)~\cite{PhysRevResearch.qsn, PhysRevA.qsn, mpe_2018, Eldredge_2018}.

\para{Initial State Optimization.}
Quantum sensing protocols typically involve four steps~\cite{RevModPhys.quantumsensing}: {\em initialization} of the quantum sensor to a desired initial state, transformation of the sensor's state over a {\em sensing} period, {\em measurement}, and {\em classical processing}.
Quantum sensor networks would have similar protocols.
In general, the initial state of the QSN can have a strong bearing on the
sensing protocol's overall performance (i.e., accuracy). E.g., in certain
settings, an entangled initial state is known to offer better estimation than
a non-entangled state~\cite{mpe_2018,Eldredge_2018}. 
If entanglement helps, then different 
entangled states may yield different estimation accuracy.
Thus, in general, determining the initial state that offers optimal estimation 
accuracy is essential to designing an optimal sensing protocol.
The focus of our work is to address this problem of determining an optimal initial 
state. Since an optimal initial state depends on the sensing and 
measurement protocol specifics, we consider a specific and concrete 
setting in this paper involving detectors.  
To the best of our knowledge, 
ours is the only work (including our recent work~\cite{Hillery_2023}) to 
address the problem of determining provably optimal initial states in quantum sensor 
networks \magentaa{with discrete outcome/parameters}.\footnote{\magentaa{For estimation of continuous parameters, 
some works~\cite{Eldredge_2018,saleem_dickestate} exist that have shown that 
certain initial states can saturate the quantum Cramer-Rao bound (also see \S\ref{sec:related}).}}

\para{QSNs with Detector Sensors.}
We consider a network of quantum ``detector'' sensors. Here,
a detector sensor is a qubit sensor whose state changes to a unique final state when an event happens. 
More formally, a sensor with initial state $\ket{\psi}$ gets transformed \magentaa{to} $U\ket{\psi}$ when an event happens, where $U$ is a particular unitary matrix that signifies the impact of an event on the sensor.  Such detector sensors can be very useful in {\em detecting} an event, rather than {\em measuring} a continuous parameter representing an environmental phenomenon. 
More generally, we 
consider a network of quantum detector sensors wherein, when an event happens, exactly one of the sensors fires--- i.e., changes its state as above.
In general, a network of such detector sensors can be deployed to {\em localize} an event --- by determining the firing sensor and, hence, the location closest to the event.
Our paper addresses the problem of optimizing the initial global state of such QSNs
to minimize the probability of error in determining the firing sensor.

\para{Contributions.}
In the above context, we make the following contributions. 
We formulate the problem of initial state optimization in detector quantum sensor networks. 
We derive necessary and sufficient conditions for the existence of an initial state that can detect the firing sensor with perfect accuracy, i.e., with zero probability of error.
Using the insights from this result, we derive a conjectured optimal solution for the problem and provide a pathway to proving the conjecture. 
We also develop multiple search-based heuristics for the problem and  
empirically validate the conjectured solution 
through extensive simulations. 
Finally, we extend our results to the unambiguous discrimination measurement scheme, non-uniform prior, and considering quantum noise.

\section{\bf \iso Problem and Related Work}
\label{sec:problem}

\begin{figure}[ht]
    \centering
    \includegraphics[width=0.85\textwidth]{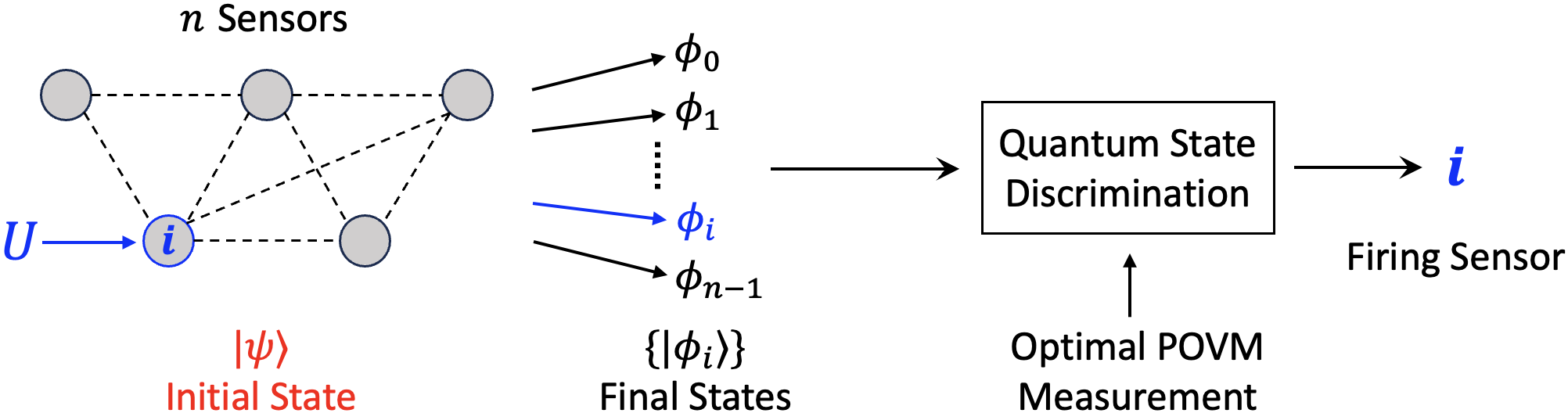}
    \caption{\iso Problem. Given $n$ deployed quantum sensors, an event changes the state of one of the sensors  ($i^{th}$ sensor in the figure) by a unitary operator $U$. Quantum state discrimination with the optimal measurement is used to determine the firing sensor. The \iso problem is to determine the initial state (possibly entangled) that minimizes the probability of error in discriminating the potential final states. The dashed lines connecting the sensors signify a potential entangled global state.
    } 
    \label{fig:qsn}
\end{figure}

\para{Setting.} Consider $n$ quantum sensors deployed across a geographical area, forming a quantum sensor network. See Fig.~\ref{fig:qsn}. 
Each sensor stores a qubit whose state may potentially change due to an event in the environment.
Let $\ket{\psi}$ denote the initial (possibly entangled) state of the $n$ sensors.
Let $U$ be a unitary operator that represents the impact of an event over a qubit in a sensor; \magentaa{here, $U$ may describe the rotation of a spin caused by a magnetic field or a phase shift induced in a state of light by a transparent object.}
Let the two eigenvectors 
of $U$ be $\{u_+, u_-\}$, and without loss of generality,
let the corresponding eigenvalues be $\{e^{+i\theta}, e^{-i\theta}\}$
where $\theta \in (0, 180)$ degrees;
thus, $U\ket{u_{\pm}}=e^{\pm i\theta}\ket{u_{\pm}}$.
Let $\ket{\phi_i} = (I^{\otimes i} \otimes U \otimes I^{\otimes (n-i-1)})\ket{\psi}$, 
where $U$ appears in the $i^{th}$ 
position and $i\in \{0, \cdots, n-1 \}$, 
represents the system's state after the event affects the $i^{th}$ sensor. 
{\em We assume that events in our setting affect exactly one sensor with uniform probability.}\footnote{\bluee{In essence, we assume that sensors are sparsely deployed such that
an event affects at most one sensor, and that the event itself is uniformly likely to occur at the
sensor locations. 
If there is no prior information about the event's location, then assuming uniform probability is reasonable.
See \S\ref{sec:extend}, where we consider the generalization of non-uniform probabilities.}} 
We refer to the $n$ possible resulting states $\{\ket{\phi_i}\}$ as the {\bf final states}; these
final states have an equal prior probability of occurrence on an event.

\softpara{Objective Function $P(\ket{\psi}, U)$.}  
When an event results in the system going to a particular final state $\ket{\phi_i}$, 
we want to determine the sensor (i.e., the index $i$) that is impacted by the event by performing a global measurement of the system. For a given setting (i.e.,  $\ket{\psi}$ and $U$), let 
$M(\ket{\psi}, U)$ be the optimal 
{\em positive operator-valued measure} (POVM) measurement for discriminating the
final states $\{\ket{\phi_i}\}$, i.e., 
the measurement that incurs the
minimum probability of error in 
discriminating the
final states $\{\ket{\phi_i}\}$ and thus
determining the index 
$i$ given an unknown final state.
Let $P(\ket{\psi}, U)$ be the (minimum) probability of error incurred by $M(\ket{\psi}, U)$ in discriminating the final states $\{\ket{\phi_i}\}$.

\para{\iso Problem Formulation.} 
Given a number of sensors $n$ and a 
unitary operator $U$, 
the \iso problem to determine the 
initial state $\ket{\psi}$ that minimizes
$P(\ket{\psi}, U)$.
In other words, we wish to determine the initial state $\ket{\psi}$ 
that yields the lowest probability of error in discriminating the final states when
an optimal POVM is used for discrimination.

{\em For clarity of presentation, we consider only the minimum error measurement 
scheme till the last Section~\ref{subsec:ud} where we extend our results to the unambiguous discrimination measurement scheme.}

\softpara{Potential Applications.}
\magentaa{One of the main applications of detector sensor networks is event localization. 
Assume we have some critical locations to monitor, and we place one quantum detector at each critical location.
Then, a network of quantum detectors, wherein a detector's
state changes (as represented by the unitary $U$), can be used to localize the event occurrence---as the location of the firing detector also gives the event's location. 
The event in the above scenario could be anything that can be represented by a unitary $U$, e.g., an event may represent the presence of a magnetic field, an acoustic event (e.g., an explosion), a signal transmission that can be detected, or movement of a detectable object.}


\para{Paper Organization.}
The rest of the paper is organized as follows. We end this section with a discussion on related work.
In the following section (\S\ref{sec:three-ortho}), 
we establish a necessary and sufficient condition for \emph{three} final states to be orthogonal---and hence, the existence of an initial state such that $P(\ket{\psi}, U) = 0$. 
We generalize the result to an \emph{arbitrary} number of sensors in \S\ref{sec:n-ortho},
and give an optimal solution for the \iso problem when the orthogonality condition is satisfied.
In \S\ref{sec:optimal}, we use the insights from \S\ref{sec:n-ortho} to derive a conjectured optimal solution for an arbitrary $U$ and the number of sensors; in the section, we also provide a pathway to proving the conjecture. 
In the following sections, we develop search-based heuristics for the problem (\S\ref{sec:searching}) and use these heuristics to 
empirically validate our conjectured solution through simulations (\S\ref{sec:sim}).
\bluee{In \S\ref{sec:extend}, we consider generalizations related to 
unambiguous discrimination measurement, non-uniform prior probabilities, and quantum noise.}
Finally in \S\ref{sec:conclusion}, we conclude and discuss some potential future work.

\subsection{Related Work}
\label{sec:related}

\para{Continuous Parameter Estimation using Quantum Sensors.}
In prior works~\cite{Eldredge_2018,Rubio_2020}, protocols have been studied for estimation of a single parameter using multiple sensors~\cite{Giovannetti_2011}, multiple parameters~\cite{mpe_2018,mpe_2020}, a single linear function over parameters~\cite{Eldredge_2018,mpe_2018,Sidhu_2020}, and multiple linear functions~\cite{Rubio_2020,Altenburg_2019}. 
Quantum state estimation considering nuisance parameters is reviewed in~\cite{Suzuki_2020}.
These and many other works~\cite{Eldredge_2018,mpe_2018,Jacobs_2018} have also
investigated whether the entanglement of sensors offers any advantage 
in improving the estimation accuracy. Some of the above works have optimized
the measurement protocols (e.g., ~\cite{Eldredge_2018,PhysRevA.qsn}) in the addressed
settings, but none of the above works have addressed the problem of initial state optimization.
To the best of our knowledge, all prior works have modeled the 
sensed parameters in the continuous domain, e.g., these parameters could be the 
strength of a magnetic field at different locations. In contrast, in some sense, 
our work focuses on estimating discrete-valued parameters. 


\para{Optimal State Discrimination.}
There has been a lot of work on quantum state discrimination~\cite{bergou-review-2007,bergou2004,Bae_2015,Barnett-review} -- wherein the goal is to determine the 
optimal measurement protocol to minimize the probability of error
in discriminating a set of given states. 
A closed-form expression is known only for two states and very specialized cases for a larger
number of states. However, numerical techniques exist (e.g., SDP-based~\cite {Eldar_2003}).
Our work differs in the following ways: (i) The set of final states we want to discriminate is very 
specialized. (ii) Our goal is to optimize the initial state---that minimizes the probability of error using an optimal POVM (in some sense, we implicitly assume that an optimal POVM for a given set of final states can be derived).



\para{Initial State Optimization.} 
Recent works have used variational circuits to seek an optimized probe 
state for a set of sensors, in the context of classical supervised learning~\cite{slaen}
and (continuous) parameter estimation~\cite{Koczor_2020} under noise. 
\eat{\cite{ouyang22-symmetric} shows that symmetric probe states are good candidates for robust quantum metrology.}
\magentaa{In additional, a recent work~\cite{ehrenberg} investigates estimation accuracy
with different levels of entanglements 
for measuring a linear combination of field amplitudes.}
In contrast, we seek provably optimal initial state solutions. 
To the best of our knowledge, the only other work that has 
investigated the initial state optimization problem is 
our recent preliminary work~\cite{Hillery_2023} where we address the same 
problem as in this paper. 
In~\cite{Hillery_2023}, we give an optimal solution for the case of $n=2$ sensors,
and, for the general case of  $n$ sensor, we derive 
close-form expressions for the probability of error 
for a heuristic solution for a restricted class of initial 
states, and investigate the benefit of entanglement in the
initial state. 

\section{\bf Orthogonality of Final States for Three Sensors}
\label{sec:three-ortho}

Note that an optimal solution for two sensors (i.e., $n=2$) is known and is based on geometric ideas (See~\cite{Hillery_2023} and~\S\ref{sec:optimal}); however,
the solution for two sensors doesn't generalize to higher $n$.
For $n \geq 3$, instead of directly determining the optimal solution, we first focus on determining the
the conditions (on $U$) under which the optimal initial state yields orthogonal final states. 
We start with the specific case of $n=3$,
as this gives us sufficient insight to generalize the results to 
an arbitrary number of sensors. 
Determining the conditions for orthogonality also helps us in conjecturing the optimal initial state for general settings. 

The basic idea for deriving the condition on $U$ that yields orthogonal final states (i.e., the below theorem) is to represent the final states on an orthonormal basis based on $U$'s eigenvalues and eigenvectors; this allows us to derive expressions for the pairwise inner products of the final states, and equating these products to zero yields the desired conditions. We now state the main theorem and proof for three sensors.

\begin{thm-prf}
Consider the \iso problem, with the unitary operator $U$, initial state $\ket{\psi}$,
and final states $\{\ket{\phi_i}\}$ as defined therein. Recall that the eigenvalues of
$U$ are $\{e^{+i\theta}, e^{-i\theta}\}$. When the number of sensors $n$ is three, \magentaa{the following is true.}

For any $\theta \in [60, 120]$ degrees, there exists a $\ket{\psi}$ such that $\ket{\phi_0}, \ket{\phi_1}, \ket{\phi_2}$ are mutually orthogonal.
Also, the converse is true, i.e., for $\theta \in (0, 60) \cup (120, 180)$, there is
no initial state that makes $\ket{\phi_0}, \ket{\phi_1}, \ket{\phi_2}$ mutually orthogonal.
\label{thm:3sensor}
\end{thm-prf}

\begin{prf}
Let us first start analyzing the inner product of $\ket{\phi_0}$ and $\ket{\phi_1}$. 
Let $z_0=\bra{\phi_0}\ket{\phi_1}$. We see that:
\begin{align*}
    z_0 & = \bra{\psi}(U^{\dagger} \otimes I \otimes I) (I \otimes U \otimes I)\ket{\psi} \\
        & = \bra{\psi}(U^{\dagger} \otimes U \otimes I) \ket{\psi}    
\end{align*}
Since $U$ is unitary, its eigenvectors $u_{-}$ and $u_{+}$ are orthogonal.
It is easy to confirm that the following eight eigenvectors of the middle-part $(U^{\dagger} \otimes U \otimes I)$ form an {\em orthonormal basis}: $\{\ket{u_{-}u_{-}u_{-}}, \ket{u_{-}u_{-}u_{+}}, \ket{u_{-}u_{+}u_{-}}, \ket{u_{-}u_{+}u_{+}}, \ket{u_{+}u_{-}u_{-}},$  $\ket{u_{+}u_{-}u_{+}}, \ket{u_{+}u_{+}u_{-}}, \ket{u_{+}u_{+}u_{+}}\}$. 
We denote these eight eigenvectors as $\{ \ket{j}  |\ j=0,\cdots,7\}$, with the $\ket{j}$ eigenvector ``mimicking'' the number $j$'s binary representation when $u_{-}$ and $u_{+}$ are looked upon as 0 and 1 respectively (so, $\ket{3}$ is 
$\ket{u_{-}u_{+}u_{+}}$).

We can write the initial state $\ket{\psi}$ in the  $\{ \ket{j} \}$ basis as
$$\ket{\psi} = \sum\limits_j \psi_j \ket{j}.$$ 
Thus, we get 
\begin{align*}
    z_0 &=  \bra{\psi}(U^{\dagger} \otimes U \otimes I) \sum\limits_j \psi_j \ket{j}   \\
        &=  \sum\limits_j |\psi_j|^{2} e_j
\end{align*} 
\noindent
where $\{e_0, e_1, \ldots, e_7\}$ are the eigenvalues corresponding to the eight eigenvectors $\{\ket{j}\}$.
\magentaa{As the eigenvalues are $1, 1, e^{+2i\theta}, e^{+2i\theta}, e^{-2i\theta}, e^{-2i\theta}, 1, 1$, we 
get:}
\begin{align}
        z_0 &= (|\psi_2|^{2} + |\psi_3|^{2})e^{+2i\theta} + (|\psi_4|^{2} + |\psi_5|^{2})e^{-2i\theta} + (|\psi_0|^{2} + |\psi_1|^{2} + |\psi_6|^{2}+ |\psi_7|^{2}) 
\end{align}
Similarly, for $z_1 = \bra{\phi_1}\ket{\phi_2} = \bra{\psi}(I  \otimes U^{\dagger} \otimes U) \ket{\psi}$, we get the below. \magentaa{Note that, in the expression for $z_1$, 
the order of eigenvalues corresponding to the coefficients 
$|\psi_i|^{2}$ is $1, e^{+2i\theta}, e^{-2i\theta}, 1, 1, e^{+2i\theta}, e^{-2i\theta}, 1$} (see Observation~\ref{obs:rhs} in \S\ref{sec:n-ortho}). Thus, we get: 
\begin{align}
    z_1 &= (|\psi_1|^{2} + |\psi_5|^{2})e^{+2i\theta} + (|\psi_2|^{2} + |\psi_6|^{2})e^{-2i\theta} + (|\psi_0|^{2} + |\psi_3|^{2} + |\psi_4|^{2}+ |\psi_7|^{2})
\end{align}
Similarly, for $z_2 =  \bra{\phi_0}\ket{\phi_2} = \bra{\psi}( U^{\dagger}  \otimes I \otimes U) \ket{\psi}$, 
we get:
\begin{align}
    z_2 &= (|\psi_1|^{2} + |\psi_3|^{2})e^{+2i\theta} + (|\psi_4|^{2} + |\psi_6|^{2})e^{-2i\theta} + (|\psi_0|^{2} + |\psi_2|^{2} + |\psi_5|^{2}+ |\psi_7|^{2})
\end{align}

Now, for $\ket{\phi_0}, \ket{\phi_1}, \ket{\phi_2}$ to be mutually orthogonal, we need $z_0 = z_1 = z_2 = 0$. This yields the following seven Equations~\ref{eqn:real1}-\ref{equ:3sensor-sum}.

\softpara{{\tt Imaginary} Equations.}
For the imaginary parts of $z_0, z_1, z_2$ to be zero, we need the following to be true. We refer to these equations as the \texttt{\textbf{Imaginary}} equations.
\begin{align}
    |\psi_2|^{2} + |\psi_3|^{2} &= |\psi_4|^{2} + |\psi_5|^{2} \label{eqn:real1}\\
    |\psi_1|^{2} + |\psi_5|^{2} &= |\psi_2|^{2} + |\psi_6|^{2}\\
    |\psi_1|^{2} + |\psi_3|^{2} &= |\psi_4|^{2} + |\psi_6|^{2} \label{eqn:real3}
\end{align}

\softpara{{\tt Real} Equations.} 
For the real parts of $z_0, z_1, z_2$ to be zero, we need the following to be true. We refer to these equations as the \texttt{\textbf{Real}} equations.
\begin{align}
   -(|\psi_2|^{2} + |\psi_3|^{2} + |\psi_4|^{2}+ |\psi_5|^{2})\cos(2\theta) &=   |\psi_0|^{2} + |\psi_1|^{2} + |\psi_6|^{2}+ |\psi_7|^{2} \label{eqn:img1} \\
   -(|\psi_1|^{2} + |\psi_5|^{2} + |\psi_2|^{2}+ |\psi_6|^{2})\cos(2\theta) &=   |\psi_0|^{2} + |\psi_3|^{2} + |\psi_4|^{2}+ |\psi_7|^{2} \\
   -(|\psi_1|^{2} + |\psi_3|^{2} + |\psi_4|^{2}+ |\psi_6|^{2})\cos(2\theta) &=   |\psi_0|^{2} + |\psi_2|^{2} + |\psi_5|^{2}+ |\psi_7|^{2} \label{eqn:img3}
\end{align}
Above, the terms with $cos(2\theta)$ are on the left-hand side (LHS), and the remaining
terms \magentaa{are} on the right-hand side (RHS).

\noindent
Finally, as $\psi_j$ are coefficients of $\ket{\psi}$, we also have
\begin{align}
\sum_j |\psi_j|^{2} &= 1  \label{equ:3sensor-sum}
\end{align}

\medskip
\noindent
{\bf Existence of $\ket{\psi}$ when $\theta \in [60, 120]$ \magentaa{that yields mutually orthogonal final states.}} 
Let us assume 
$|\psi_0|^2 = |\psi_7|^2 = y$ and $|\psi_i|^2 = x$ for $1 \leq i \leq 6$. 
These satisfy Equations~\ref{eqn:real1}-\ref{eqn:real3}, and the Equations~\ref{eqn:img1}-\ref{eqn:img3} yield: 
\begin{align}
    -4x\cos(2\theta) &= 2x + 2y  \nonumber \\
    -(2\cos(2\theta) + 1)x &= y  \nonumber
\end{align}
The above has a valid solution (i.e., $x, y \geq 0$, and $2y + 6x = 1$ from
Eqn.~\ref{equ:3sensor-sum}) when 
$\cos(2\theta) \leq -\frac{1}{2}$ i.e., when $\theta \in [60, 120]$.

\medskip
\noindent
{\bf When $\theta \in (0, 60) \cup (120, 180)$, no existence of $\ket{\psi}$  \magentaa{that yields mutually orthogonal final states.}}
Let $a = |\psi_0|^{2} + |\psi_7|^{2}$. Then, by using Equation~\ref{eqn:real1} in
Equation~\ref{eqn:img1} and so on, we get the following:
\begin{align*}
    -2 (|\psi_4|^{2} + |\psi_5|^{2})\cos(2\theta) &= a + |\psi_1|^{2} + |\psi_6|^{2} \\
    -2 (|\psi_2|^{2} + |\psi_6|^{2})\cos(2\theta) &= a + |\psi_3|^{2} + |\psi_4|^{2} \\
    -2 (|\psi_1|^{2} + |\psi_3|^{2})\cos(2\theta) &= a + |\psi_2|^{2} + |\psi_5|^{2} 
\end{align*}
Adding up the above equations and rearranging, we get:
\begin{align*}
 (-2\cos(2\theta) - 1) \sum_{j=1}^6 |\psi_j|^{2} &= 3a
\end{align*}
Thus, we need $(-2\cos(2\theta) - 1) \geq 0$, as $a \geq 0$, i.e., we need
$\cos(2\theta) \leq -\frac{1}{2}$. 
Thus, for $\theta \in (0, 60) \cup (120, 180)$, 
there is no solution to the above equations.  {\em Note that we have not used any symmetry argument here.}
\end{prf}

\section{\bf Orthogonality of Final States for $n$ Sensors}
\label{sec:n-ortho}

In this section, we generalize the result in the previous section to an arbitrary number of sensors greater than 3.\footnote{For two sensors, the single equation corresponding to the Equations~\ref{eqn:img1}-\ref{eqn:img3} can be made equal to zero on both sides with $\theta=45$ degrees and zeroing all coefficients on the RHS (which is possible due to lack of other equations).}

\begin{theorem}
Consider the \iso problem, with the unitary operator $U$, initial state $\ket{\psi}$,
and final states $\{\ket{\phi_i}\}$ as defined therein. Recall that the eigenvalues of
$U$ are $\{e^{+i\theta}, e^{-i\theta}\}$. Let $n \geq 3$ be the number of sensors. \magentaa{The following is true.} 

For any $\theta \in [T, 180-T]$ degrees, there exists a $\ket{\psi}$ such that the set of $n$ states $\{\ket{\phi_i}\}$ are mutually orthogonal, where $T$ is given by:
$$T = \frac{1}{2}\arccos{\left(-\frac{\lceil \frac{n}{2} \rceil - 1}{\lceil \frac{n}{2} \rceil}\right)}$$
Note that $T \in (45, 90)$ degrees. In particular, the values of $T$ for increasing $n$ are:  60 ($n=4)$, 65.9 ($n=5,6$), 69.3 ($n=7,8$), 71.6 ($n=9,10$).

\magentaa{The converse of the above} is also true, i.e., 
for $\theta \in (0, T) \cup (180-T, 180)$, there is
no initial state $\ket{\psi}$  that makes $\{\ket{\phi_i}\}$  mutually orthogonal. 
\label{thm:nsensors}
\end{theorem}
    

\medskip
Before we prove the theorem, we define the partitioning of coefficients and state an observation.

\para{Partitioning the Coefficient-Squares $\{|\psi_j|^2\}$ into ``Symmetric'' Sets.} 
Note that just renumbering the sensors does not change the optimization problem. 
Based on this intuition, we can group the eigenvectors $\ket{j}$
(and the corresponding coefficients $\psi_j$'s) into equivalent classes. 
Let $n$ be the number of sensors.  
Since only the coefficient-squares $\{|\psi_j|^2\}$ appear in the expression for pairwise inner-products of the final states, 
we just partition the coefficient-squares rather than the coefficients $\{\psi_j\}$ themselves---as only the coefficient-squares are relevant to our proposed solution and discussion. 
We partition the set of $2^n$ coefficient-squares into $n+1$ symmetric sets $\{S_k\}$ as follows: 
$$S_k =   \{|\psi_j|^{2}\ |\ \ket{j}\ {\rm has}\ k\  {\rm number\ of}\ u_{+}\}\ \ \forall \ 0 \leq k \leq n$$

\noindent
For each $0 \leq k \leq n$, let $R_k$ be the number of coefficient-squares from $S_k$ in the RHS of a {\tt Real} equation, and $L_k$ be the number of coefficient-squares from $S_l$ in the LHS of  {\tt Real} equation. (Note that, by Observation~\ref{obs:rhs} below, for any $k$, the number of coefficient-squares of $S_k$ that are in the RHS (LHS) is the same for all {\tt Real} equations.)
{\bf For the case of} $\mathbf{n=3}$, we have
$S_0 = \{|\psi_0|^{2}\}, 
S_1 = \{|\psi_1|^{2},  |\psi_2|^{2},  |\psi_4|^{2} \}, 
S_2 = \{|\psi_3|^{2},  |\psi_5|^{2},  |\psi_6|^{2} \}, 
S_3 = \{|\psi_7|^{2} \}.$ Also, we have $R_0 = R_1 = R_2 = R_3 = 1$, while $L_0 = L_3 = 0$, $L_1 = L_2 = 2$. 
We will use the above terms to prove the theorem. 
\medskip
\medskip

\begin{observation}
For a {\tt Real} equation $E$ corresponding to the inner-product of final states $\phi_i$ and $\phi_j$ (for $0 \leq i, j \leq n-1$), a coefficient-square $|\psi_r|^2$ appears in the RHS of the equation $E$ iff the bit-representation of the number $r$ has either both 0's or both 1's at the $i^{th}$ and $j^{th}$ most-significant bits.
\label{obs:rhs}
\end{observation}

%

\begin{lem-prf}
For $n\geq3$,
$$min_{1 \leq k \leq (n-1)} \frac{R_k}{L_k} =  \frac{\lceil \frac{n}{2} \rceil - 1}{\lceil \frac{n}{2} \rceil}.$$ 
Thus, for the given $T$ in Theorem~\ref{thm:nsensors}, $L_k\cos(2T) + R_k = 0$ for some $k$, and 
$R_k + \cos(2T)L_k \geq 0$ for all $k$.
\label{lemma:t}
\end{lem-prf}
\begin{prf}
For  $n \geq 3$ and $0 \leq k \leq n$, from Observation~\ref{obs:rhs} we get that: 
$$R_k = {n-2 \choose k-2} + {n-2 \choose k}$$
$$L_k = 2 {n-2 \choose k-1}.$$
Above, we assume ${x \choose y} = 0$ if $y > x$ or $y < 0$.
Now, a simple analysis shows that:
$$\left(min_{2 \leq k \leq (n-2)} \frac{{n-2 \choose k-2} + {n-2 \choose k}}{2 {n-2 \choose k-1}}\right) =  \frac{\lceil \frac{n}{2} \rceil - 1}{\lceil \frac{n}{2} \rceil}$$ 
Since, for $n \geq 3$, $R_1 = R_{n-1} = n-2$ and $L_1=L_{n-1}=2$, we get the lemma. 
\end{prf}

\begin{observation}
Let $\sum_i x_i =c$, for a set of variables $x_i \geq 0$ and a constant $c > 0$.
The equation $\sum_i c_ix_i = 0$, where $c_i$ are some constants, has a solution if and only if (i) at least one of the constants is positive and one of the constants is negative, or (ii) one of the constants is zero.
\label{obs:pos-neg}
\end{observation}

\subsection{\bf Proof of Theorem~\ref{thm:nsensors}.}

\begin{prf} 
\softpara{If $\theta \in [T, 180-T]$.} Let the set of all coefficient-squares in each $S_k$ to be equal to 
$x_k$, for each $k$. Then, each {\tt Imaginary} equation becomes:
\begin{equation}
\sum_{k=0}^{n}  (L_k/2)x_k  = \sum_{k=0}^{n} (L_k/2)x_k \label{eqn:imag-final}
\end{equation}
Each {\tt Real} equation becomes:
\begin{equation}
    -\cos(2\theta) \sum_{k=0}^{n}    L_kx_k  = \sum_{k=0}^{n} R_k x_k 
    \label{eqn:real}
\end{equation}
\begin{equation}
  \sum_{k=0}^{n} ( R_k + \cos(2\theta) L_k) x_k = 0 \label{eqn:real-final}  
\end{equation}

By Observation~\ref{obs:pos-neg}, the above equation (and thus, all {\tt Real} equations) can be made true by appropriate choices of $x_k$ since 
\begin{enumerate}
    \item $R_k + \cos(2\theta)L_k$ is positive for $k=0$ as $L_0 = 0$ and $R_0 = 1$.
    \item $R_k + \cos(2\theta)L_k$ is negative or zero for some $k$ by Lemma~\ref{lemma:t} when $\theta \in [T, 180-T]$.
\end{enumerate}

\softpara{\magentaa{If $\theta \in (0, T) \cup (180-T, 180)$}.}
Adding all the {\tt Real} equations gives the following. Below, $f(j) = k$ such that $|\psi_j|^2 \in S_k$.

$$ -\cos(2\theta)\sum_{j=0}^{2^n} {n \choose 2} \frac{L_{f(j)}}{|S_{f(j)}|} |\psi_j|^2	= \sum_{j=0}^{2^n} {n \choose 2} \frac{R_{f(j)}}{|S_{f(j)}|}  |\psi_j|^2 $$
The above gives: 
$$\sum_{j=0}^{2^n} \frac{1}{|S_{f(j)}|} (R_{f(j)} + \cos(2\theta)L_{f(j)})  |\psi_j|^2 = 0$$
The above equation doesn't have a solution as $(R_k + \cos(2\theta)L_k) > 0$ for all $k$
for \magentaa{$\theta \in (0, T)$ (and thus, for $\theta \in (180-T, 180)$)} for by Lemma~\ref{lemma:t}.
\end{prf}

\para{Optimal Initial State under Theorem~\ref{thm:nsensors}'s Condition.}
Based on the above theorem, we can derive the optimal initial state under the condition of Theorem~~\ref{thm:nsensors}; the optimal initial state yields mutually-orthogonal final states.

\begin{cor-prf}
\label{cor:orthogonal-opt}
Consider the \iso problem, with the unitary operator $U$, initial state $\ket{\psi}$,
and final states $\{\ket{\phi_i}\}$ as defined therein. Recall that the eigenvalues of
$U$ are $\{e^{+i\theta}, e^{-i\theta}\}$. Let $n \geq 3$ be the number of sensors.
When $\theta \in [T, 180-T]$ degrees, where $T$ is defined in Theorem~\ref{thm:nsensors}, an 
optimal initial state $\ket{\psi}$ that yields
mutually orthogonal final states $n$ states $\{\ket{\phi_i}\}$ is given as follows.\footnote{We note that there are many optimal solutions.}

Let $S_l$ be the partition that minimizes the ratio $R_l/L_l$. It follows from Lemma~\ref{lemma:t}'s proof (we omit the details) that $l = \lfloor \frac{n}{2} \rfloor$, and $R_l, L_l,$ and $S_l$ are given by:
\begin{align*}
L_l &= |S_l| \times \frac{\lceil \frac{n}{2} \rceil}{2\lceil \frac{n}{2} \rceil-1}\\
R_l &= |S_l| \times \frac{\lceil \frac{n}{2} \rceil - 1}{2\lceil \frac{n}{2} \rceil-1} \\
|S_l| &= {n \choose \lfloor \frac{n}{2} \rfloor  }
\end{align*}
Then, the coefficients of an optimal initial state $\ket{\psi}$, 
when $\theta \in [T, 180-T]$ degrees with $T$ defined in Theorem~\ref{thm:nsensors}, 
are such that their coefficient-squares are as follows:
\begin{align*}
|\psi_j|^{2} &= \frac{1}{|S_l| - \cos(2\theta) L_l - R_l} \ \hspace{0.2in} \forall\ |\psi_j|^{2} \in S_l  \\
|\psi_j|^{2} &= \frac{-\cos(2\theta)L_l - R_l}{|S_l| - \cos(2\theta) L_l - R_l}   \ \hspace{0.17in}   \ \forall\ |\psi_j|^{2} \in {S_0} \\
|\psi_j|^{2} &= 0\ \hspace{1.19in}   \ \forall\ |\psi_j|^{2} \notin {S_l \cup S_0}
\end{align*}
\label{thm:optimal-largerT}
\end{cor-prf}

\begin{prf}
The proof of the above Corollary easily follows from the fact that each 
coefficient-square of the solution is positive (from Lemma~\ref{lemma:t}), and that
the coefficient-squares of the solution satisfy Eqn.~\ref{eqn:real-final} (and Eqn.~\ref{eqn:imag-final} trivially) as well as the constraint in Eqn.~\ref{equ:3sensor-sum}.
\end{prf}
\section{\bf Conjectured Optimal \iso Solution}
\label{sec:optimal}

\para{Provably Optimal Solution for Two Sensors.}
The above joint-optimization problem for the case of 2 sensors can be solved optimally as follows. 
First, we note that the minimum probability of error in discriminating two final states for a given initial state $|\psi\rangle$  is given by: 

\begin{equation}
P_{e} = \frac{1}{2} \left( 1 - \sqrt{ 1 - |\langle \psi |(U\otimes U^{-1} ) |\psi\rangle |^{2}} \right). 
\label{eqn:two}
\end{equation}

Now, when the eigenvalues of 
$U$ are $\{e^{+i\theta}, e^{-i\theta}\}$, as in our \iso problem, then
the initial state $|\psi\rangle$ that minimizes the above
probability of error for $0 \leq \theta \leq \pi /4$ and 
$3\pi/4 \leq \theta \leq \pi$
can be shown to be the following entangled state:
 \begin{equation}
 |\psi\rangle= \frac{1}{\sqrt{2}} (|u_{+}\rangle |u_{-}\rangle + |u_{-}\rangle |u_{+}\rangle ).
 \label{eqn:two-sol}
 \end{equation}
For 
$\pi /4 \leq \theta \leq 3\pi/4$, there exists an initial state that
yields orthogonal final states. 
The above follows from the techniques developed to 
distinguish between two unitary operators~\cite{DAriano_2002};
we refer the reader to our recent work~\cite{Hillery_2023} for
more details. Unfortunately, the 
above technique doesn't generalize to $n$ greater than 2, because
for greater $n$, there is no corresponding closed-form expression for minimum
probability of error.

\para{Conjectured Optimal Solution For $n$ Sensors.}
The main basis for our conjectured optimal solution is that an 
optimal initial state
must satisfy the {\em symmetry of coefficients} property which is defined as
follows: an initial state satisfies the {\em symmetry of coefficients} property, if for each $k$, the set of coefficient-squares in $S_k$ have the same value.
The {\em intuition} behind why an optimal initial state must satisfy the 
{\em symmetry of coefficients} property comes from the following facts: 
\begin{enumerate}
\item 
The \magentaa{optimal initial state}, under the condition of Theorem~\ref{thm:nsensors}, satisfies the symmetry of coefficients property. 

\item
Since sensors are homogeneous, ``renumbering'' the 
sensors doesn't change the optimization problem 
instance fundamentally. 
Thus, if $\ket{\psi}$ is an optimal initial state, then all initial
state solutions obtained by permuting the orthonormal 
basis $\{\ket{j}\}$ corresponding to a renumbering of 
sensors,\footnote{Note that renumbering the sensors is tantamount to
renumbering the bits in the bit-representation of $j$ integers
of the orthonormal basis $\{\ket{j}\}$. 
See Theorem~\ref{thm:final}'s proof for more details.}
must also yield optimal initial 
states.\footnote{Note that this fact doesn't imply that the optimal solution must satisfy the 
symmetry of coefficients property.}
Now, observe that an initial state that satisfies
the symmetry of coefficients property remains unchanged under 
any permutation of the orthonormal 
basis $\{\ket{j}\}$ corresponding to a renumbering of sensors.

\item 
Similarly, due to the homogeneity of sensors, an optimal initial state 
must yield ``symmetric'' final states---i.e., final 
state vectors that have the same pairwise angle between them.
Now, from Observation~\ref{obs:rhs}, we observe that
an initial state that satisfies the symmetry of coefficients yields
final states such that their pairwise inner-product value is the same.
\end{enumerate}
Finally, it seems intuitive that this common (see \#3 above) 
inner-product value of every pair of final 
states should be minimized to minimize the probability of error in discriminating the final states. Minimizing the common inner-product value within the problem's constraints yields the below optimal solution conjecture.


\medskip
\begin{conjecture}
\label{conj:opt}
Consider the \iso problem, with the unitary operator $U$, initial state $\ket{\psi}$,
and final states $\{\ket{\phi_i}\}$ as defined therein. Recall that the eigenvalues of
$U$ are $\{e^{+i\theta}, e^{-i\theta}\}$. Let $n \geq 3$ be the number of sensors.
For a given $\theta \in (0, T] \cup [180-T, 180),$ degrees, where $T$ is from Theorem~\ref{thm:nsensors}, the optimal initial state $\ket{\psi}$ for the \iso 
problem is as follows. 

Let $S_l$ be the partition 
that minimizes $(R_l + \cos(2\theta)L_l)/(R_l + L_l)$,
where $R_l$ and $L_l$ are as defined in the previous section. 
The coefficients of the optimal solution are such that their coefficient-squares are given by:
\begin{align*}
|\psi_j|^{2} &= 1/|S_l|\ \ \ \ \forall\ |\psi_j|^{2} \in S_l  \\
|\psi_j|^{2} &= 0\ \hspace{0.3in}   \ \forall\ |\psi_j|^{2} \notin S_l
\end{align*}
\label{thm:optimal}
\end{conjecture}
We note the following: (i) The above conjecture optimal solution is provably optimal for $n=2$, with $T = 45$ degrees; see Eqn.~\ref{eqn:two-sol} above and~\cite{Hillery_2023}.
(ii) The above conjectured optimal solution yields orthogonal final states for $\theta = T$. 
In particular, it can be easily shown that the above conjectured optimal solution
is the same as the solution derived in Corollary~\ref{cor:orthogonal-opt} for $\theta = T$.
(iii) The proposed state in the above conjecture is a Dicke State in the basis made up of $\ket{u_{-}}$ and $\ket{u_{+}}$.
Dicke states can be prepared deterministically by linear depth quantum circuits in a single quantum computer~\cite{dicke_state}, \magentaa{and be prepared in a distributed quantum network as well~\cite{dickestate_distributed}}.
We now show that the above conjecture can be proved with the help of the following simpler conjecture.


\para{Proving Symmetry of Coefficients.} Based on the intuition behind the above Conjecture~\ref{conj:opt}, one way to prove it would be to prove the symmetry of coefficients---i.e., the existence of an optimal solution wherein the coefficient-squares in any $S_k$ are equal. Proving symmetry of coefficients directly seems very challenging, but we believe that the below conjecture (which implies symmetry of coefficients, as shown in Theorem~\ref{thm:final}) is likely more tractable. Also, {\em the below Conjecture has been verified to hold true in our empirical study (see \S\ref{sec:sim})}. 

\begin{conjecture}
For a given $U$,
consider two initial states 
$\ket{\psi} = \sum\limits_j \psi_j \ket{j}$ and 
$\ket{\psi'} = \sum\limits_j \psi'_j \ket{j}$
such that (i) they are two unequal coefficient-squares, i.e., for some $j$, $|\psi_j|^2 \neq |\psi'_j|^2$, 
and 
(ii) they have the same objective function value, i.e., $P(\ket{\psi}, U) = P(\ket{\psi'}, U)$. 
We claim that the ``average'' state given by 
$$\ket{\psi_{avg}} = \sum_j \sqrt{\frac{|\psi_j|^2 + |\psi'_j|^2}{2}}  \ket{j}$$ 
has a lower objective function value, i.e., $P(\ket{\psi_{avg}}, U) < P(\ket{\psi'}, U)$.
\label{conj:avg}
\end{conjecture}

We now show that the above Conjecture is sufficient to prove the optimal solution Conjecture~\ref{conj:opt}.

\begin{thm-prf}
Conjecture~\ref{conj:avg} implies Conjecture~\ref{conj:opt}. 
\label{thm:final}
\end{thm-prf}

\begin{prf}
We start by showing that Conjecture~\ref{conj:avg} implies 
the symmetry of coefficients, and then minimize the common pairwise inner-product values of
the final states.

\softpara{Conjecture~\ref{conj:avg} implies \magentaa{Symmetry} of Coefficients.}
First, note that for a given initial state $\ket{\psi}$, we can generate ($n!-1)$ 
other ``equivalent'' initial states (not necessarily all different) by 
just renumbering the sensor (or, in other words, permuting the basis eigenvectors).
Each of these initial states should yield the same objective value $P()$ 
as that of $\ket{\psi}$, as it can be shown that they would 
yield essentially the same set of final states.
As an example, the following two initial states are  
\magentaa{equivalent} (i.e., \magentaa{yield} the same objective value $P()$); 
here, the sensors numbered 1 and 2 have been interchanged. 
$$\psi_0 \ket{0} +  \psi_1 \ket{1} + \psi_2 \ket{2} + \psi_3 \ket{3} +
\psi_4 \ket{4} + \psi_5 \ket{5} + \psi_6 \ket{6} + \psi_7 \ket{7}$$
$$\psi_0 \ket{0} +  \psi_2 \ket{1} + \psi_1 \ket{2} + \psi_3 \ket{3} +
\psi_4 \ket{4} + \psi_6 \ket{5} + \psi_5 \ket{6} + \psi_7 \ket{7}$$
More formally, for a given initial state $\ket{\psi} =  \sum_j \phi_j \ket{j}$, 
a permutation (renumbering of sensors) $\pi: \{0, 1, \ldots, n-1\} \mapsto \{0, 1, \ldots, n-1\}$ yields
an equivalent initial state given by
$\ket{\psi'} =  \sum_j |\phi_\Pi(j)| \ket{j}$ where $\Pi: \{0, 1, \ldots, 2^n-1\} \mapsto \{0, 1, \ldots, 2^n-1\}$ is 
such that $\Pi(j) = i$ where the bits in the bit-representation of $j$ are permuted using $\pi$ to yield $i$.
It can be shown that the set of final states yielded by $\ket{\psi}$ and $\ket{\psi'}$ 
are essentially the same (modulo the permutation of basis dimensions), 
and hence, they will yield the same probability of error and thus objective value $P()$.

Now, consider an optimal initial state $\ket{\psi} = \sum_j |\phi_j| \ket{j}$ 
that doesn't have symmetry of coefficients---i.e., there is a pair of 
coefficient-squares $|\phi_i|^2$ and $|\phi_j|^2$ such that
they are in the same set $S_k$ but are not equal.
The numbers $i$ and $j$ have the same number of 1's and 0's in their binary representation, 
as $|\phi_i|^2$ and $|\phi_j|^2$ belong to the same set $S_k$.
Let  $\Pi$ be a permutation function (representing renumbering of the $n$ sensors)
such that $\Pi(i) = j$. Consider an initial state 
$\ket{\psi'} = \sum_j \psi_{\Pi(j)} \ket{j}$, which has the same probability of 
error as $\ket{\psi}$. Now, applying Conjecture~\ref{conj:avg} on $\ket{\psi}$ and 
$\ket{\psi'}$ yields a new initial state with a lower objective value $P()$, which 
contradicts the optimality of $\ket{\psi}$. Thus, all optimal initial-states 
must satisfy the symmetry of coefficients.

\softpara{Maximizing the Pairwise Angle.}
Now, an optimal initial state with symmetry of coefficients will yield 
final states that have the same pairwise inner-product values (this follows from Theorem~\ref{thm:nsensors}'s proof). 
Also, we see that each pairwise inner-product value is (see Eqns.\ref{eqn:imag-final} and \ref{eqn:real-final} from \S\ref{sec:n-ortho}):
\begin{equation}
\sum_{k=0}^{n} ( R_k + \cos(2\theta) L_k) x_k \label{eqn:common}
\end{equation}
with the constraint that
$$ \sum_{k=0}^{n} ( R_k + L_k) x_k = 1.$$

{\em When $\theta \in (0, T]$.}
By Lemma~\ref{lemma:t}, note that $(R_k + \cos(2\theta) L_k) x_k \geq 0$ for all $k$, for 
$\theta \in (0,T)$. 
We show in Lemma~\ref{lemma:angle} below that, for states with equal and positive
pairwise inner-products, the probability of error in discriminating them using an optimal
measurement increases with an increase in the common inner-product value. 
Thus, the optimal initial state must minimize the above inner-product value expression in Eqn.~\ref{eqn:common}.
Now, from Observation~\ref{obs:pos-neg2} below, the inner-product value above is minimized when the coefficient-squares in the $S_l$ that minimizes $(R_k + \cos(2\theta) L_k)/|S_l|$ are non-zero, while the coefficient-squares in all other 
$S_k$'s where $k \neq l$ are zero. This proves the theorem.

{\em When $\theta \in [180-T, 180)$.} 
Note that $\cos(2\theta) = \cos(2(180-\theta))$, and since $(180-\theta) \in (0, T]$ for  
$\theta \in [180-T, 180)$, we can use the same argument as above for this case as well.
\end{prf}

\begin{observation}
Let $\sum_i a_i x_i =1$, for a set of positive-valued variables $x_i$ and positive constants $a_i$. 
The expression $\sum_i c_ix_i$, where constants $c_i$'s are all {\em positive}, has a minimum value of $\min_i c_i/a_i$ which is achieved by $x_i = 1/a_i$ for the $i$ that minimizes $\min_i c_i/a_i$. 
\label{obs:pos-neg2}
\end{observation}

\para{Minimizing Probability of Error in Discriminating ``Symmetric'' Final States.} 
We now show, using prior results, that 
if the pairwise inner-products (and hence, angles) 
of the resulting final states $\ket{\phi_i}$ are equal, then the 
probability of error in discriminating them 
is minimized when the pairwise inner-product values are minimized. 

\begin{lem-prf}
Consider $n$ states to be discriminated $\phi_0, \phi_1, \ldots, \phi_{n-1}$ 
such that $\bra{\phi_i}\ket{\phi_j} = x$,
for all $0 \leq i, j \leq n-1$ and $i \neq j$. 
The probability of error in discriminating 
$\phi_0, \phi_1, \ldots, \phi_{n-1}$ using an optimal measurement 
increases with an increase in $x$ when $x \geq 0$.
\label{lemma:angle}
\end{lem-prf}
\begin{prf}
The optimal/minimum probability of error using the optimal POVM for a set of states with equal pairwise inner products can be computed to be~\cite{quantum_pyramid}:
$$P_e = 1- \frac{1}{n}\left(\sqrt{1-\frac{(n-1)(1-x)}{n}} + (n-1)\sqrt{\frac{1-x}{n}}\right)^2$$
Let the inner term be $y$, such that $P_e = 1 - (y^2/n)$. The derivative of $y$ with respect to $x$ is given by:
$$ \frac{n-1}{2\sqrt{n}}\left(\frac{1}{\sqrt{nx+1-x}} - \frac{1}{\sqrt{1-x}}\right).$$
The above is negative for $x \geq 0$. Thus, for a given number of sensors $n$ and $x \geq 0$, the probability of error $P_e$ increases with an increase in $x$.
\end{prf}

\para{Summary.} 
In summary, we propose the Conjecture~\ref{conj:opt} for the optimal solution for the \iso problem, based on the symmetry of coefficients. 
We also propose a Conjecture~\ref{conj:avg}, which seems more straightforward to prove and provably implies Conjecture~\ref{conj:opt}. 
We empirically validate these conjectures using several search heuristics in the following sections.  
\section{Search Heuristics}
\label{sec:searching}

\begin{algorithm}[h] 
  	\KwIn{The initial state $\vx$, $i$th element of $\vx$, step size}
	\KwOut{A neighbor $\vx'$ of $\vx$}
	$\vx' \leftarrow \vx$ \;
	$direction \leftarrow$ Generate a random unit 2D-vector \;
	$direction' \leftarrow$ convert $direction$ to complex number \;
	$\vx'[i] \leftarrow \vx'[i] + direction' \times stepSize$ \;
	$\vx' \leftarrow Normalize(\vx'$) \tcp*{$\vx^{\dagger}\vx = 1$}
	\Return $\vx'$ \;
	\caption{FindNeighbor($\vx, i, stepSize$)}
\label{algo:find-neighbor}
\end{algorithm}

In this section, we design three search heuristics to determine an efficient
\iso solution, viz., hill-climbing algorithm, simulated annealing, and genetic algorithm. In the next section, we will evaluate these heuristics and observe
that they likely deliver near-optimal solutions. 
We start with a numerical (SDP) formulation of determining an optimal 
measurement, and thus, develop a method to estimate the objective value
$P(\ket{\psi}, U)$ of a given initial state $\ket{\psi}$.

\para{Semi-Definite Program (SDP) for State Discrimination.}
We now formulate a 
semi-definite program (SDP) to compute the optimal measurement 
for a given initial state; this formulation allows us to determine the optimal measurement 
using numerical methods, and thus, facilitates the development of the
search heuristics for the \iso problem as described below.
Given a set of (final) quantum states, determining the optimal measurement that
discriminates them with minimum probability of error is a convex optimization problem, and in particular, can be formulated as a semi-definite program~\cite{Eldar_2003}.
Let the final states be $\{\ket{\phi_i}\}$ with
prior 
probabilities $p_{i}$, where $\sum_i p_i = 1$.
Let $\{\Pi_{i}\}$ be the POVM operator with $\Pi_{i}$ being the element associated with the state $\ket{\phi_i}$, and let $Tr()$ be the trace operator.
The SDP program to determine the optimal measurement operator can be formulated
as below, where the objective is to minimize the probability of error.
\begin{equation}
    \min_{\Pi_{i} \in \calb} 1 - \sum_{i=0}^{n-1} p_{i} Tr(\Pi_{i} \ket{\phi_i}\bra{\phi_i} )
    \label{eqn:measure-sdp}
\end{equation}
subject to the constraints:
\begin{align}
    \Pi_{i} & \succeq 0, \ \ \ \   0 \leq i \leq n-1 \label{eqn:semidefinite} \\
    \sum_{i=0}^{n-1} \Pi_{i} &= I \label{eqn:identity}
\end{align}
Above, Eqn.~\ref{eqn:semidefinite} ensures that every measurement operator is positive semidefinite, while Eqn.~\ref{eqn:identity} ensures that the set of measurement operators is a valid POVM, i.e., the summation of all measurement operators is the identity matrix.
Eqn.~\ref{eqn:measure-sdp} minimizes the probability of error expression for a given POVM measurement and set of quantum states.

\begin{algorithm}[t] 
  	\KwIn{Unitary operator $U$}
        \KwIn{Number of sensor $n$}
	\KwOut{Initial State $\vx$}
	$\vx \leftarrow$ a random state vector with a length of $2^{n}$\;
	$bestObjective \leftarrow P(\vx, U)$ \;
        $stepSize \leftarrow$ 0.1 \;
        $stepDecreaseRate \leftarrow$ 0.96\ \;
	\While{Termination Condition Not Satisfied}{
	    \For{$i = 1$ to  $2^{n}$}{
	        $neighbors \leftarrow$ Find 4 neighbors, call FindNeighbor($\vx, i, stepSize$) four times \;
	        $bestStep \leftarrow 0$ \;
	        \For{$j = 1$ to $4$}{
	            $objective \leftarrow P(neighbors[j], U)$  \;
	            \If{$objective < bestObjective$}{
	                $bestObjective \leftarrow objective$ \;
	                $bestStep \leftarrow j$      \;
	            }
	        }
    	    \If{$bestStep$ is not 0}{
    	        $\vx \leftarrow neighbors[bestStep]$ \;
    	    }
	    }
	    $stepSize \leftarrow stepSize\times stepDecreaseRate$ \;
	}
	\Return $\vx$ \;
	\caption{HillClimbing($U$, $n$)}
\label{algo:hill-climbing}
\end{algorithm}

\para{The Objective Value of an Initial State.}
To design the search-based heuristics, we need a method to estimate an objective value for a given initial quantum state that evaluates its quality.
In our context, for a given initial state $\ket{\psi}$, 
the \iso problem's objective function $P(\ket{\psi}, U)$ could
also serve as the objective function in a search-based heuristic.
$P(\ket{\psi}, U)$ can be directly estimated
using the Eqn.~\ref{eqn:measure-sdp} above.
\begin{equation}
P(\ket{\psi}, U) = 1-\sum_{i=0}^{n-1} p_{i} Tr(\Pi_{i}  \ket{\phi_i}\bra{\phi_i})
\label{eqn:objective}
\end{equation}
\noindent
where $\ket{\phi_i} = (I^{\otimes i} \otimes U \otimes I^{\otimes (n-i-1)})\ket{\psi}$ are the final states, and the optimal measurement $\{\Pi_i\}$ can be computed numerically using the SDP formulation given above. 

Based on the above method to estimate the objective function $P()$, we can develop  search heuristics for the \iso problem; 
at a high level, each heuristic searches for a near-optimal initial state by starting with a random initial state \vx\ and iteratively improving (not necessarily in every single iteration) by moving to \vx's better neighbor based on the objective value $P()$ of \vx.

\begin{algorithm}[b] 
  	\KwIn{Unitary operator $U$}
        \KwIn{Number of sensor $n$}
	\KwOut{Initial State $\vx$}
	$\vx \leftarrow$ a random state vector with a length of $2^{n}$\;
	$stepSize \leftarrow 0.1$ \;
	$T \leftarrow$ Standard deviation of some \vx \ neighbors' objective values \;
    $stepDecreaseRate \leftarrow$ 0.96 \;
    $coolingRate \leftarrow$ 0.96\ \;
    $stdRatio \leftarrow 1$ \;
	\While{Termination Condition Not Satisfied}{
	    \For{$i = 1$ to $2^{n}$}{
                \For{$j=1$ to $4$}{
                    $\vx' \leftarrow$ FindNeighbor($\vx, i, stepSize$) \;
	            $E_1 \leftarrow P(\vx, U)$  \;
	            $E_2 \leftarrow P(\vx', U)$  \;
    	        $\Delta E \leftarrow E_2 - E_1$ \;
    	        \uIf{$\Delta E < 0$}{
    	             $\vx \leftarrow \vx'$
    	        }
    	        \Else{
    	            $\vx \leftarrow \vx'$ with probability $e^{-\Delta E/T}$
    	        }
                }
	    }
	    $stepSize \leftarrow stepSize \times stepDecreaseRate$ \;
	    $std \leftarrow$ Standard dev. of \vx\ recent neighbors' scores\;
	    $stdRatio \leftarrow stdRatio\times coolingRatio$ \;
	    $T \leftarrow min(T\times coolingRate, std\times stdRatio)$ \;
	}
	\Return $\vx$ \;
	\caption{SimulatedAnnealing($U$, $n$)}
\label{algo:simulated-annealing}
\end{algorithm}

\para{Hill-Climbing\footnote{In our context of a minimization problem, the heuristic actually {\em descends} into a valley of solutions, but we stick to the Hill-Climbing name because that's the common usage.} (HC) Search Heuristic.}
The Hill-climbing (HC) heuristic starts with randomly picking an initial quantum state for the $n$-sensors, i.e., a $2^{n}$ length vector $\vx$ of complex numbers with $\vx^{\dagger}\vx = 1$.
During each iteration, we look into one element of the state vector $\vx$ at a time.
And for each element, we look into four random ``neighbors'' of the initial state (as described below), 
and pick the neighbor with the lowest objective value $P()$. 
We repeat the process until reaching the termination criteria, i.e., the improvement (if any) of moving to the best neighbor is smaller than a threshold (i.e., $10^{-6}$). 
We also set a minimum number of 100 iterations.

To find a neighbor of a quantum state, 
we update one element of the state vector \vx\ at a time---by 
adding to it a random unit vector multiplied by a step size
which decreases with each iteration (a post-normalization 
step is done to maintain $\vx^{\dagger}\vx = 1$).
For each element, we look into four random neighbors instead of one, to increase the chance of discovering better neighbors.
See Algo.~\ref{algo:find-neighbor} for the neighbor-finding procedure and Algo.~\ref{algo:hill-climbing} for the overall Hill Climbing heuristic.

\para{Simulated Annealing (SA) Heuristic.}
The above Hill-climbing heuristic can get stuck in a local optimal.
Thus, we also develop a more sophisticated Simulated Annealing 
(SA)~\cite{Kirkpatrick_1983} metaheuristic which has a mechanism to 
jump out of a local minimum.
By convention, SA applies the concept of energy $E$.
In our context, the energy is the equivalent of the objective function value $P()$.
In essence, SA allows itself to transition to solutions with worse objective values with a small (but non-zero) probability.
In SA, the transition probability to a new neighbor state depends upon the improvement
$\Delta E$ in the objective function and is given by: 
\begin{equation}
    P(\Delta E) = \min (1, e^{-\Delta E/T}),
    \label{eqn:boltzmann}
\end{equation}
where $T$ is the temperature.
We note that when the new state's objective value is lower, then $\Delta E$ is negative, and thus, $P(\Delta E)$ is 1, and the new state is readily transitioned to. 
Same as in~\cite{SA-sudoku}, we set the initial temperature as the standard deviation of the objective value of several initial state's neighbors. 
As the SA algorithm iterates, the temperature $T$ gradually decreases. 
In our context, the following works well and leads to fast convergence, compared to other standard equations used in other contexts~\cite{Laarhoven_1987}.
\begin{equation}
    T_{n} = min\{(1-\epsilon)T_{n-1}, (1 - \epsilon)^{n}\sigma_{n-1}\},
    \label{eqn:our-anneal}
\end{equation}
where $\sigma_{n-1}$ is the standard deviation of objective values of the latest ten neighbors explored at the $(n-1)^{th}$ iteration.
SA uses the same neighbor-finding method (Algo.~\ref{algo:find-neighbor}) as in the previous Hill-climbing heuristic, with a similar termination condition as Hill-climbing except that we 
allow a few iterations for improvement.
The pseudo-code of SA is shown in Algo.~\ref{algo:simulated-annealing}.

\begin{algorithm}[b] 
  	\KwIn{Unitary operator $U$}
        \KwIn{Number of sensor $n$}
	\KwOut{Initial State $\vx$}
        $N \leftarrow$ population size\;
	$\vx_{pop} \leftarrow$ a size $N$ population of length $2^{n}$ random state vectors\;
	\While{Termination Condition Not Satisfied}{
            $ranks \leftarrow$ $computeRank(\vx_{pop}, U)$ \;
            $\vx_{pop}' \leftarrow$ an empty children population\;
	    \While{$length(\vx_{pop}') < size$}{
                $parents \leftarrow$ get two states by $select(ranks, \vx_{pop})$ \;
                $children \leftarrow$ get two new states by $twoPointCrossover(parents)$ \;
                Do mutation for $children$ \;
                Add  $children$ to $\vx_{pop}'$   \;
	    }
             $\vx_{pop} \leftarrow$ the top $N$ of states in $\vx_{pop} + \vx_{pop}'$ \;
        }
        $\vx \leftarrow$ the best state in $\vx_{pop}$   \;
	\Return $\vx$ \;
	\caption{GeneticAlgorithm($U$, $n$)}
\label{algo:genetic-algorithm}
\end{algorithm}

\para{Genetic Algorithm (GA) Heuristic.}
The Genetic Algorithm (GA) is another popular metaheuristic algorithm for solving optimization problems.
Inspired by the natural evolution of survival of the fittest~\cite{Holland_1992}, 
GA works by considering a ``population'' of candidate solutions and creating the
next generation iteratively, until the best solution in a new generation does not
improve from the best solution in the previous generation by at least a threshold.
In our context, the initial population of candidate solutions is a set of random initial 
states. Candidate solutions are evaluated by a fitness function, which is conceptually the same as our objective function $P()$ (Eqn.~\ref{eqn:objective}) except that the fitness function is higher the better while $P()$ is lower the better.
So, $1-P()$ will serve as the fitness function for GA.
The pseudo-code for GA is shown in Algo.~\ref{algo:genetic-algorithm}.
To create a new generation, we pick a pair of candidate states as parents through the rank selection~\cite{review-ga} and then generate a pair of children states by using the two-point crossover method~\cite{review-ga}. 
Finally, we mutate the children in a way similar to finding neighbors in Algo.~\ref{algo:find-neighbor}.

\section{Validating the Conjectures Empirically}
\label{sec:sim}

In this section, we evaluate our search heuristics for varying $U$ operator (i.e., varying values of $\theta$) and for $n = 2$ to $5$ and observe that they likely deliver
near-optimal initial state solutions to the \iso problem. 
Based on this observation, we show that our optimal solution Conjecture~\ref{conj:opt} is very
likely true as well the Conjecture~\ref{conj:avg}.
Our search heuristics implementation and 
experiment's raw data are open-source at~\cite{iso-code}.


\begin{figure}
    \centering
    \includegraphics[width=0.6\textwidth]{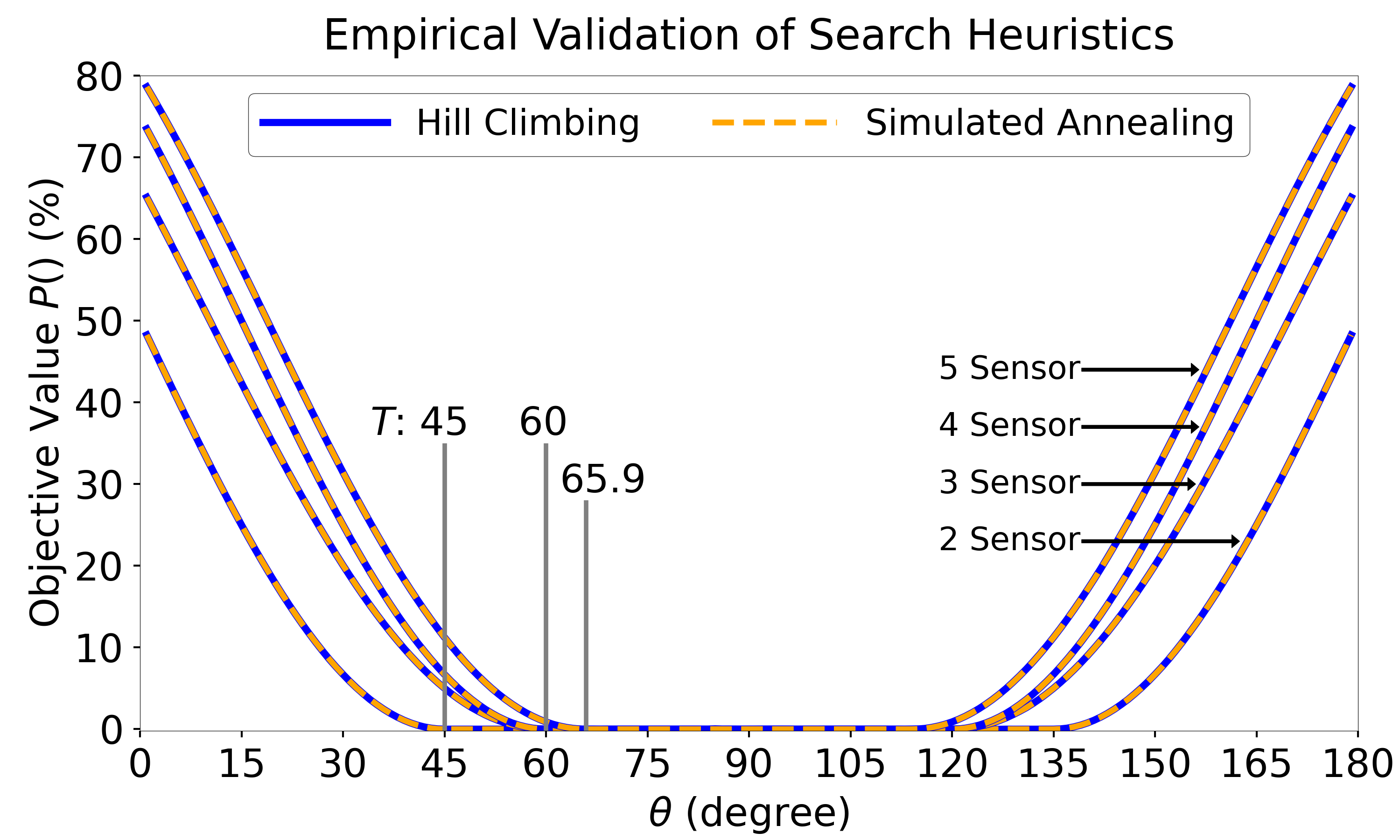}
    \vspace{-0.1in}
    \caption{Performance of the three search heuristics for varying $U$'s parameter $\theta$, for different number of sensors in the network. Genetic Algorithm (GA) is not shown explicitly, for clarity, but it also performs almost exactly the same as Hill-Climbing and Simulated Annealing (SA) which are plotted above.}
    \label{fig:heuristics}
\end{figure}

\begin{figure}
    \vspace{-0.15in}
    \centering
    \includegraphics[width=0.6\textwidth]{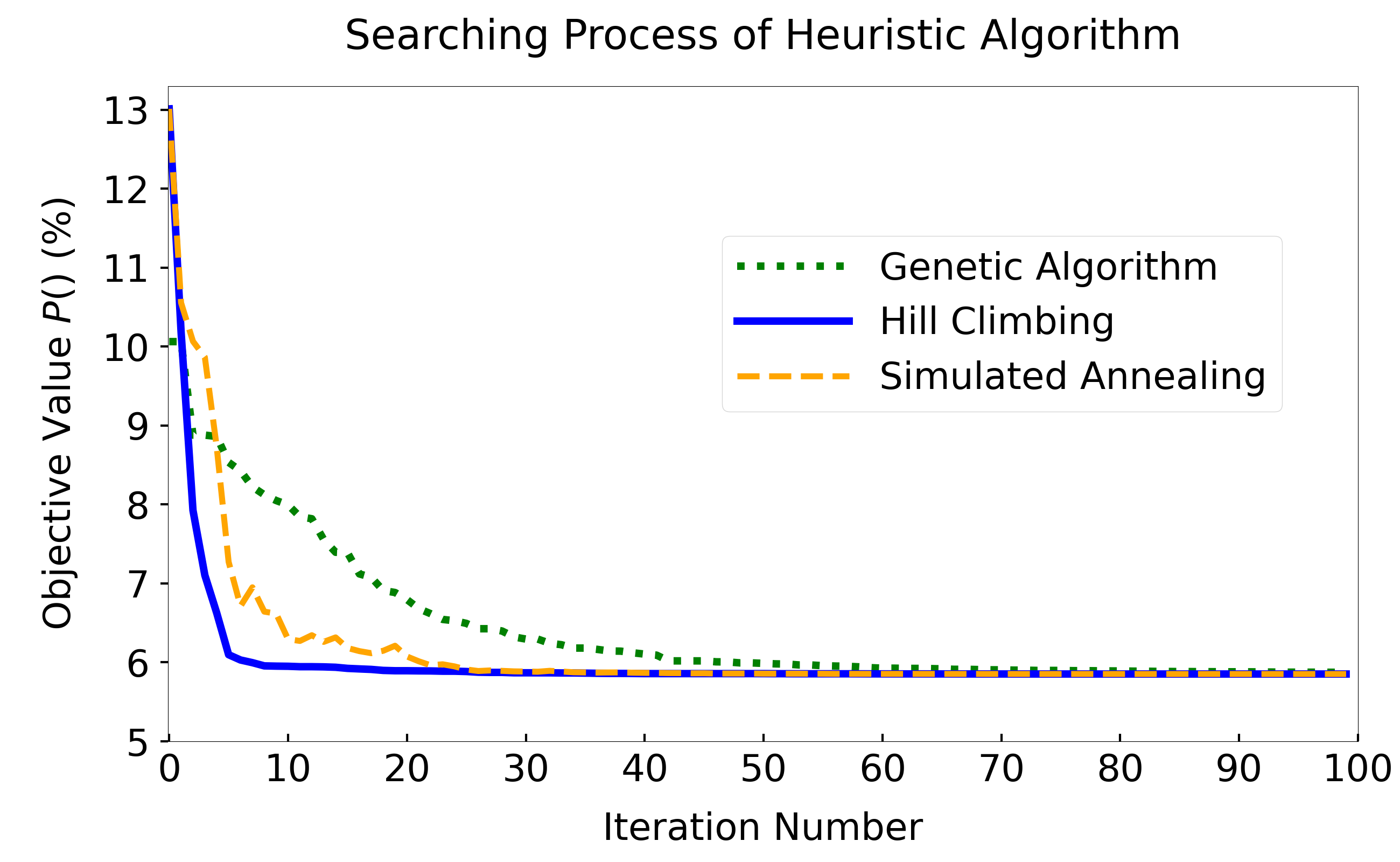}
    \caption{The objective value $P()$, probability of error, of the candidate solution over iterations of the 
    three search heuristics for a special value of $\theta =46$ degrees and $n = 4$ sensors.}
    \label{fig:iterations}
\end{figure}

\para{Evaluation Setting.}
Recall that, without loss of generality, we assume the eigenvalues of 
$U$ to be $\{e^{+i\theta}, e^{-i\theta}\}$ with $U\ket{u_{\pm}}=e^{\pm i\theta}\ket{u_{\pm}}$ where $u_{\pm}$ are the
two eigenstates of $U$.
In our evaluations, we vary the $\theta$ in the range of $(0, 180)$ degrees, and
assume the prior probabilities of final states to be uniform. We consider four values of $n$, the number of sensors, viz., 2, 3, 4, and 5. Running simulations for much larger values of $n$ is infeasible due to the prohibitive computational resources needed. 
E.g., the estimated computation time to run any of the search 
heuristics for $n=10$ will take 10s of years, based on our 
preliminary 
estimates.\footnote{In our context, the Hill-Climbing heuristic goes through about 100
iterations and in each iteration, it needs to solve $4\cdot2^n$ instances of SDP formulations 
(Eqns~\ref{eqn:measure-sdp}-\ref{eqn:identity}) where $n$ is the number of sensors. 
We use the Convex-Python CVXPY~\cite{diamond2016cvxpy} package
(which in turn used the Splitting Conic Solver~\cite{sdp-solver}) to solve our SDP formulations, and observe that it takes more than an hour to 
solve a single SDP instance for $n=10$; this suggests an estimate of 10s of years of computation time for $n=10$.}




\para{Performance of Search Heuristics.}
Fig.~\ref{fig:heuristics} shows the performance of the search heuristics
under varying $\theta$ and four values of $n = 2, 3, 4, 5$, in terms of the \iso objective function $P(\ket{\psi}, U)$ for the initial state solution $\ket{\psi}$. We make the following two observations:
\begin{enumerate}
    \item All three heuristics perform almost exactly the same.
    \item The heuristics deliver an initial state solution with $P(\ket{\psi}, U) = 0$ for the same range of $\theta$ given in Theorem~\ref{thm:nsensors}.
\end{enumerate}
We also observe that the heuristics perform the same for $\theta$ and $\pi - \theta$,
i.e., symmetric along the $\theta=\pi/2$ line. Thus, in the remaining plots, we only plot
results for $\theta \in (0, \pi/2]$.
Fig.~\ref{fig:iterations} shows the convergence rates of the three heuristics for a specific value of $\theta=46$ degrees and $n=4$ sensors. 
We observe that HC converges the fastest, followed by SA and GA.
After 100 iterations, the HC and SA end at a probability of error of $5.85\%$, while GA ends at $5.86\%$.

\begin{figure}
    \centering
    \includegraphics[width=0.6\textwidth]{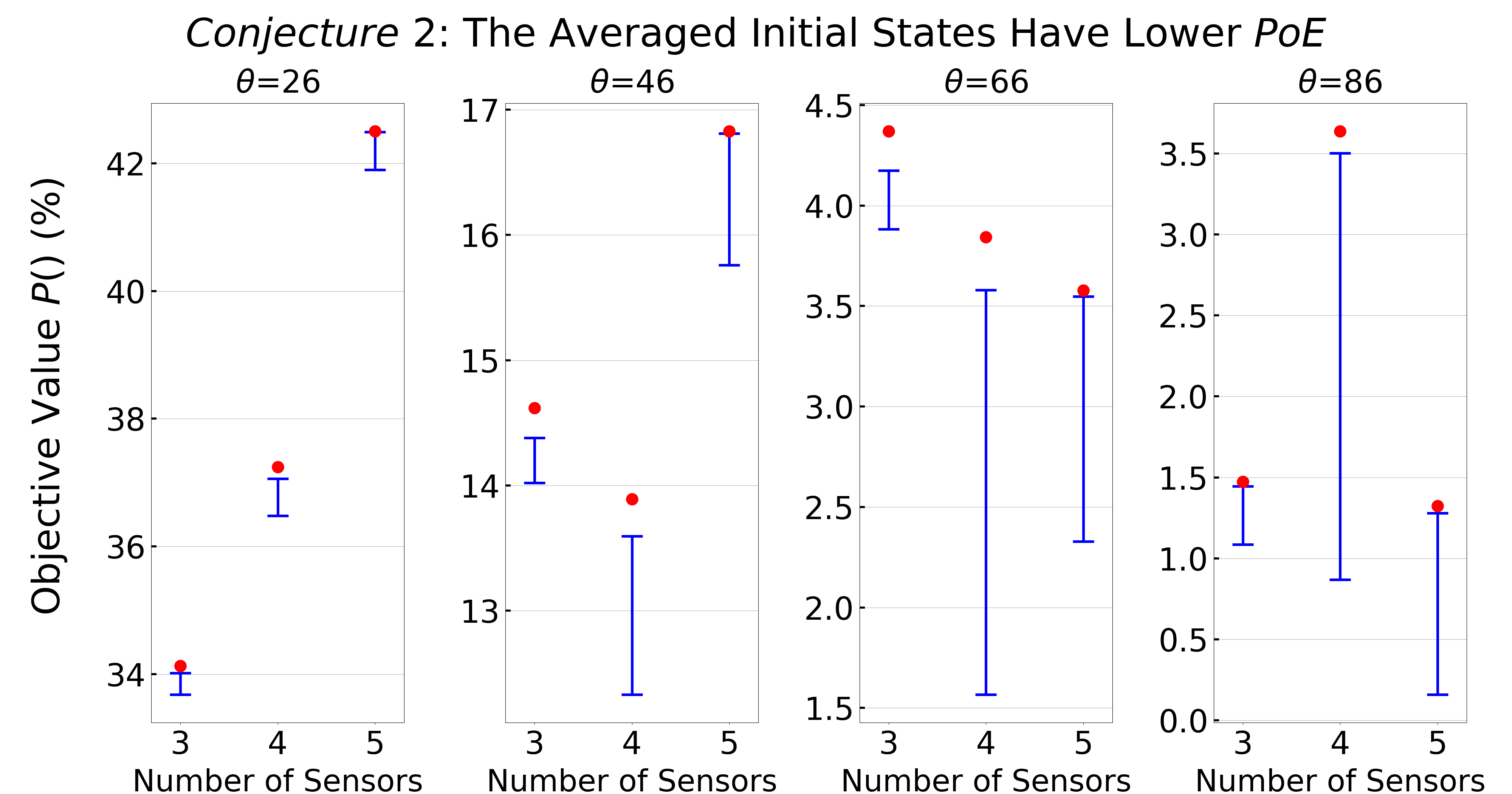}
    \caption{Empirical validation of Conjecture~\ref{conj:avg}. For four different values of $\theta$ and three different values of $n$, we show that
    the objective value (Probability of Error) of the original initial state (the red circle) remains higher than the objective value of the many ``averaged'' states (range shown by blue the bar).}
    \label{fig:lemma2}
\end{figure}

\begin{figure}
    \vspace{-0.15in}
    \centering
    \includegraphics[width=0.6\textwidth]{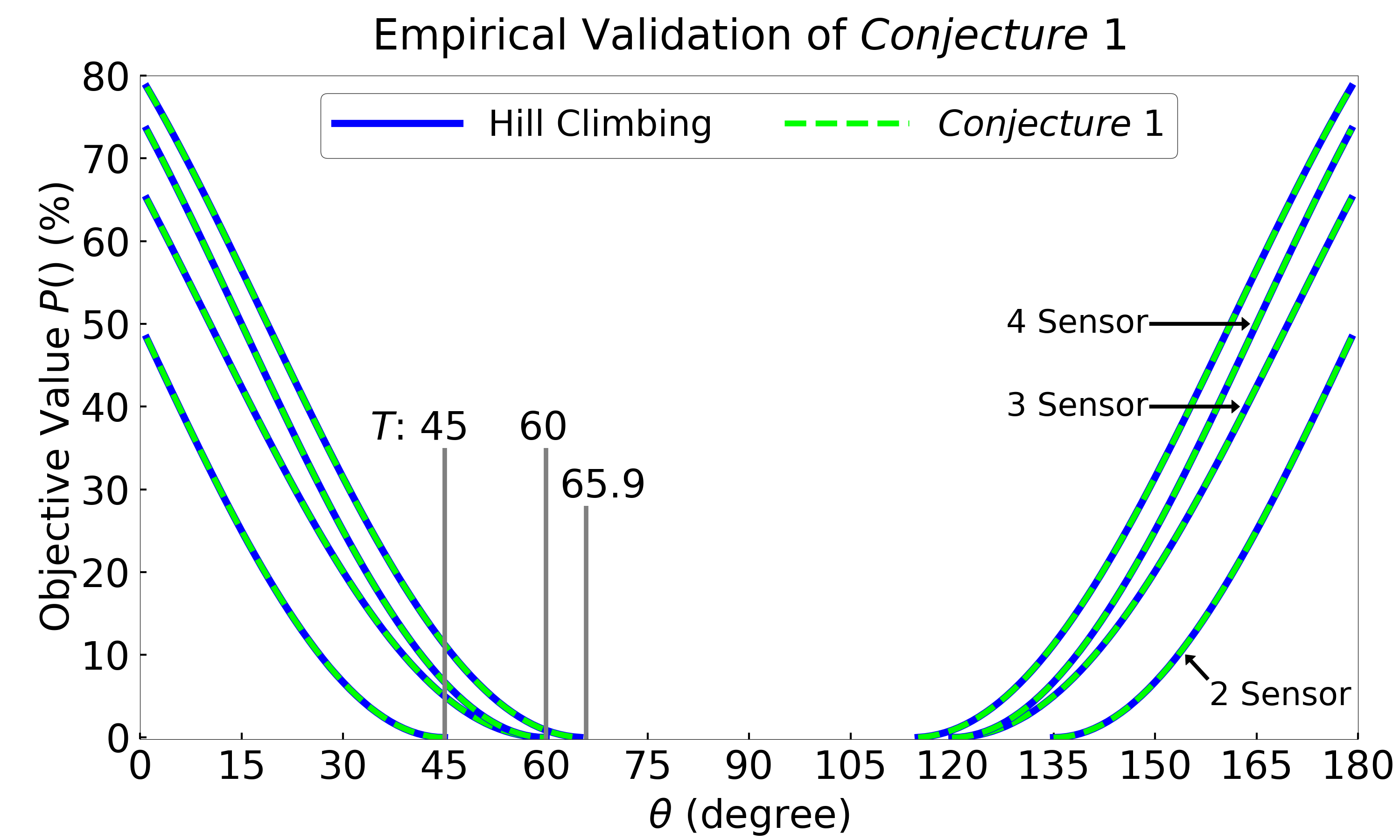}
    \caption{The Conjecture~\ref{conj:opt}'s solution performs almost exactly as the Hill-Climbing heuristic when $\theta \in (0, T] \cup [180-T, 180),$ degrees, where $T$ is from Theorem~\ref{thm:nsensors}. 
    For $n=2$, Conjecture~\ref{conj:opt}'s solution matches with the provably optimal solution from~\cite{Hillery_2023} with $T$ being 45 degrees.}
    \label{fig:conjecture}
\end{figure}

\para{Empirical Validation of Conjecture~\ref{conj:avg}.}
Recall that Conjecture~\ref{conj:avg} states that an ``average'' solution 
of two \iso solutions with
equal objective values have a lower objective value. To empirically validate
Conjecture~\ref{conj:avg}, we generate a random state $\ket{\psi}$, and then, 
generate  $n!-1$ additional states of the same objective value $P()$ by 
renumbering the sensors as discussed in Theorem~\ref{thm:final}'s proof. 
Then, we take many pairs of these states, average them, and compute the
objective value. 
Fig.~\ref{fig:lemma2} plots the objective value of the original state $\ket{\psi}$,
and the range of the objective values of the averaged states. We observe that the
objective values of the averaged states are invariably less than those of 
$\ket{\psi}$.

\para{Empirical Validation of the Optimal Solution Conjecture~\ref{conj:opt}.}
We now evaluate the performance of the initial state solution obtained by Conjecture~\ref{conj:opt} and compare it with the solution delivered by 
one of the search heuristics--Hill Climbing (HC).
Here, we consider $\theta \in (0, T) \cup (180-T, 180)$ degree, where $T$ is 
as defined in Theorem~\ref{thm:nsensors}.
In Fig.~\ref{fig:conjecture}, we observe that the HC heuristic and Conjecture~\ref{conj:opt} solutions have identical performance, suggesting that
Conjecture~\ref{conj:opt}'s solution is likely optimal based on our earlier observation
that the search heuristics likely deliver optimal solutions.



\begin{figure}
    \centering
    \begin{subfigure}[t]{0.495\textwidth}
	\includegraphics[width=\textwidth]{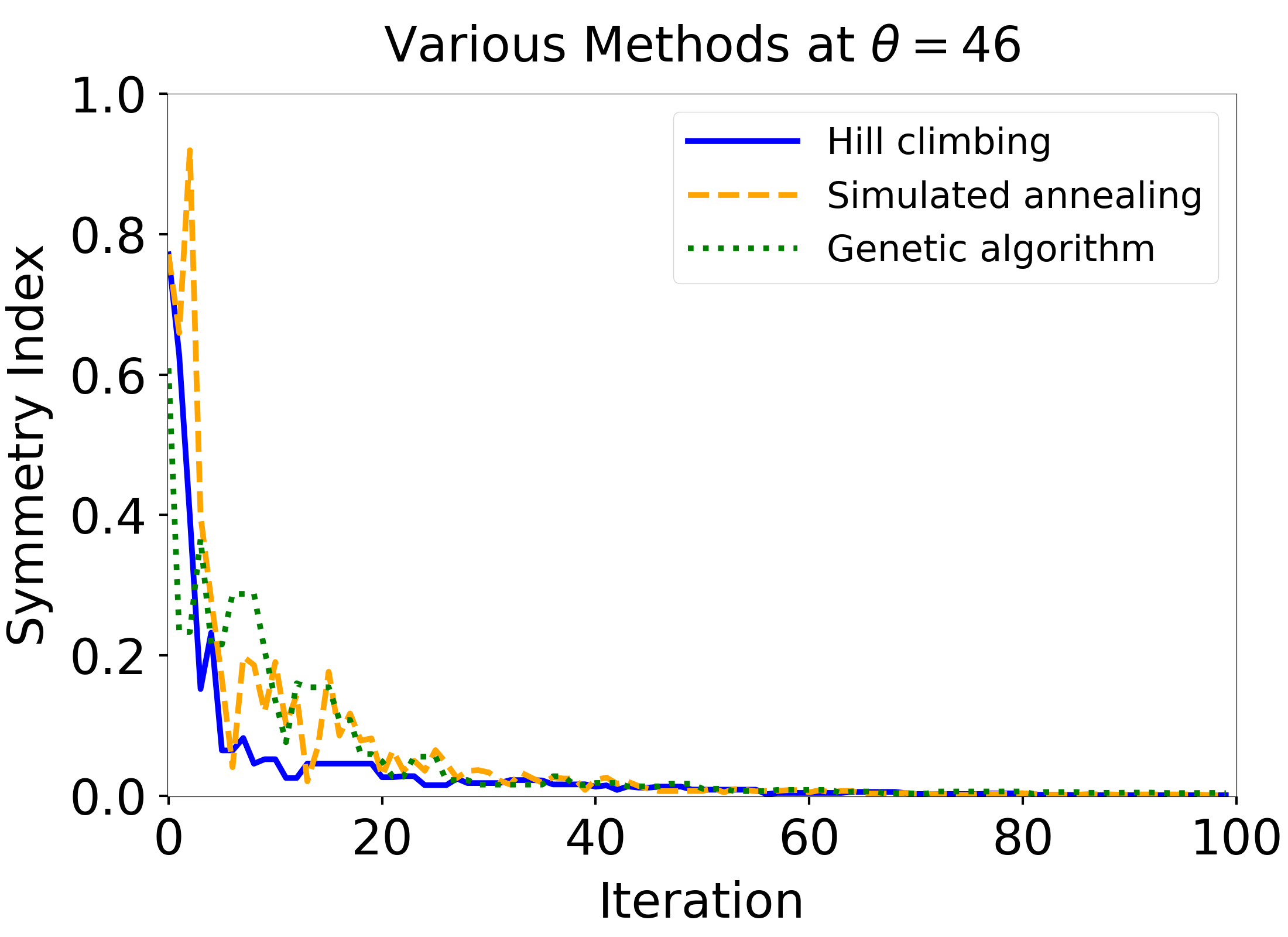}
	\caption{The three search heuristics for $\theta = $ 46 degrees. }
        \label{fig:symmetry-methods}
    \end{subfigure}
    \qquad
    \hspace{-0.3in}
    \begin{subfigure}[t]{0.495\textwidth}
	\includegraphics[width=\textwidth]{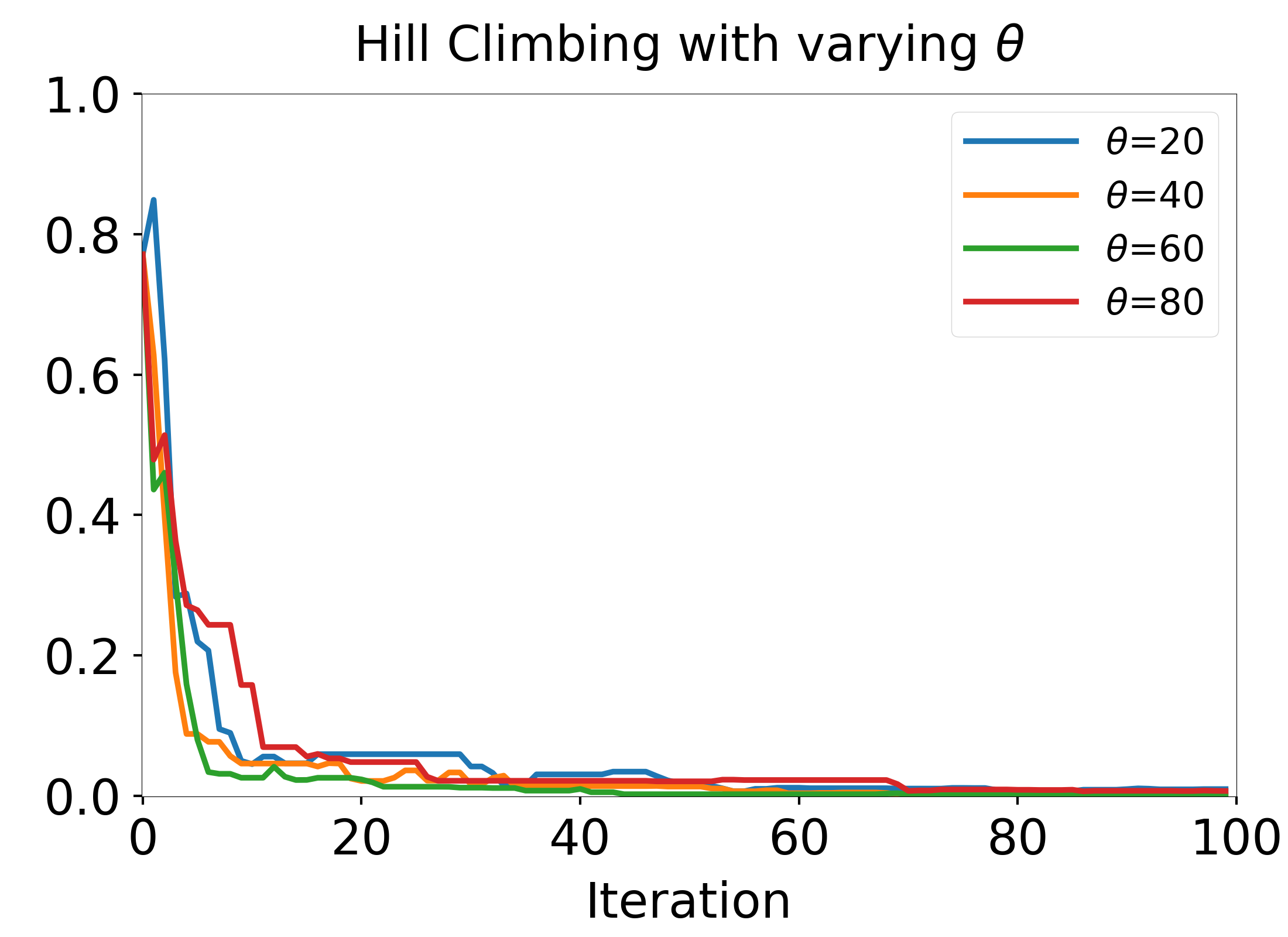}
	\caption{The HC heuristic for different values of $\theta$.}
        \label{fig:symmetry-theta}
    \end{subfigure}
    \caption{Symmetry-index of the candidate solutions over iterations.}
    \label{fig:symmetry-iteration}
\end{figure}

\begin{figure}
    \centering
    \begin{subfigure}[t]{0.495\textwidth}
	\includegraphics[width=\textwidth]{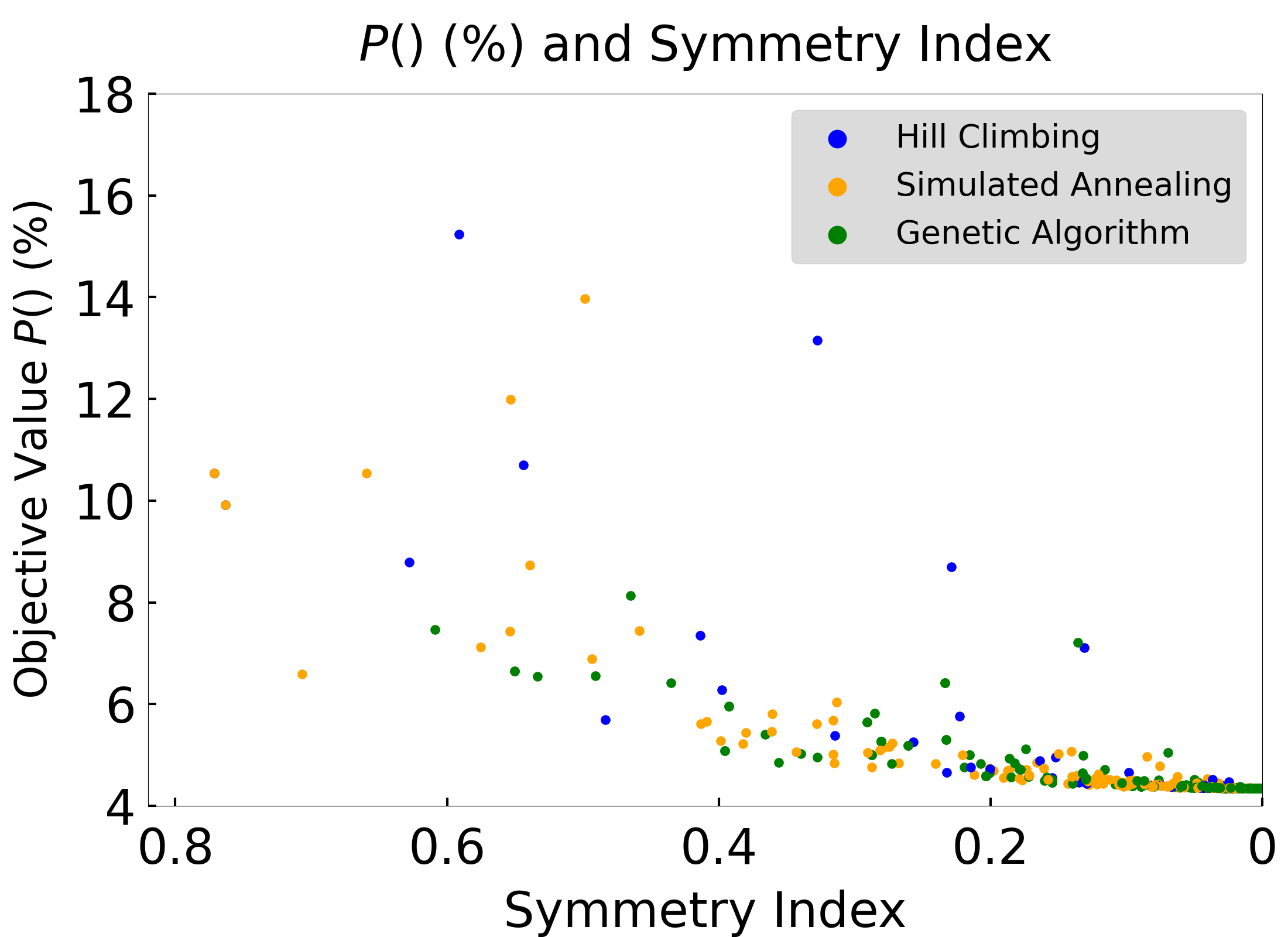}
	\caption{Probability of Error vs.\ Symmetry Index}
        \label{fig:poe_symmetry}
    \end{subfigure}
    \qquad
    \hspace{-0.3in}
    \begin{subfigure}[t]{0.495\textwidth}
	\includegraphics[width=\textwidth]{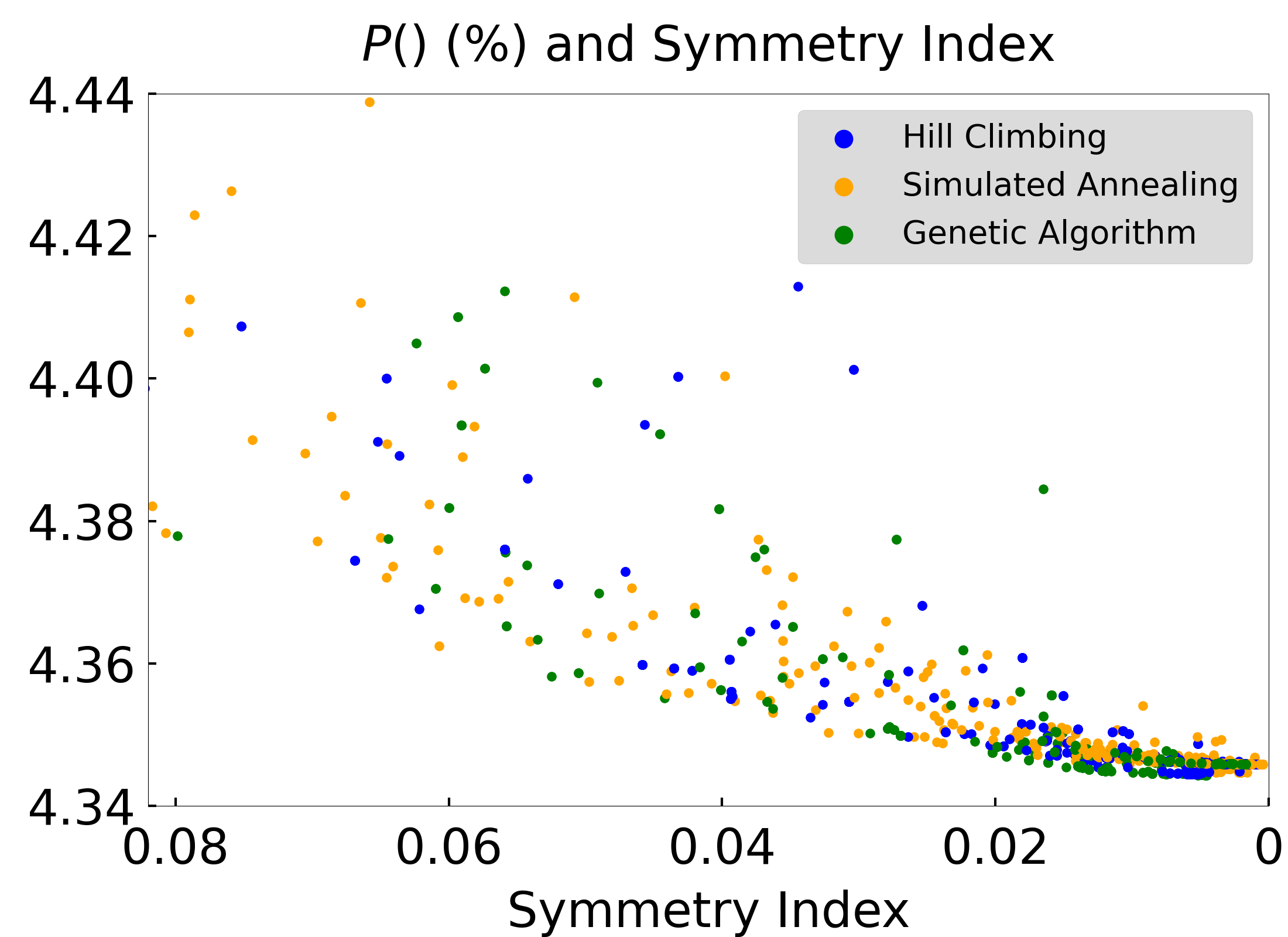}
	\caption{Probability of Error vs.\ Symmetry Index, zoomed-in for low values of symmetry index}
        \label{fig:poe_symmetry_zoom}
    \end{subfigure}
    \caption{The correlation between the objective value (probability of error) and the symmetry index.}
\end{figure}

\para{Symmetry ``Index'' vs.\ Objective Value (Probability of Error).}
In this final experiment, we investigate the impact of the symmetry of coefficients on the objective value of an initial state. 
Here, we only do experiments for $n=3$ number of sensors.
To this end, we define a notion of {\em symmetry index} which quantifies the symmetry of coefficients in a given initial
state. In particular, we define the {\em symmetry index} for an initial state
$\ket{\psi} = \sum\limits_j \psi_j \ket{j}$ as:  
\begin{equation}
\sum_{k=0}^{n} \sum_{|\psi_i|^2, |\psi_j|^2 \in S_k} (|\psi_i|^2 - |\psi_j|^2)^2
\label{eqn:symmety} 
\end{equation}
\noindent
where $S_k$ is the $k$th symmetric set as defined in Theorem~\ref{thm:nsensors}.
The symmetric index being zero implies that within each symmetric set, all the coefficient-squares are equal.
Fig.~\ref{fig:symmetry-iteration} shows that the search heuristic essentially generates solutions with lower and lower symmetry index and, finally, converges to a solution with zero symmetry index value. 
This is true for all three search heuristics (Figs.~\ref{fig:symmetry-methods}) and for varying $\theta$ (Fig.~\ref{fig:symmetry-theta}).
Given that Fig.~\ref{fig:iterations} already shows that the objective value decreases as the searching iterations go on, we can conclude that the objective value and the symmetry index decrease simultaneously when the iterations go on.
Furthermore in Fig.~\ref{fig:poe_symmetry}, we show the correlation between symmetry index and objective value through a scatter plot---with the objective value generally decreasing with the decrease in symmetry index. 
Fig.~\ref{fig:poe_symmetry_zoom} zooms into the later iterations of the heuristics wherein the symmetry index is very low (less than 0.08) to show a clearer view of the correlation.



\section{Extensions}
\label{sec:extend}

In this section, we consider the generalization to
the unambiguous discrimination scheme, \bluee{non-uniform prior probabilities,
and quantum noise.}

\subsection{Unambiguous Discrimination Measurement}
\label{subsec:ud}

Till now, we have only considered the minimum error measurement scheme wherein the measurement operator always outputs a state, though sometimes incorrectly and thus incurring a certain probability of error. 
We now consider an alternative measurement scheme of unambiguous measurement~\cite{Barnett-review} where there are no errors, but the measurement can fail, i.e. giving an inconclusive outcome.
The unambiguous measurement scheme thus may incur a probability of failure. 
Fortunately, our results for the minimum error measurement scheme also hold for the unambiguous discrimination measurement scheme and objective, as observed below.

\begin{enumerate}
\item
The sufficient and necessary condition for orthogonality derived in Theorem~\ref{thm:nsensors} is a property of the states and the operator $U$, 
and is independent of the measurement scheme. Thus, Theorem~\ref{thm:nsensors}
hold for an unambiguous discrimination scheme.

\item
The intuition behind Conjecture~\ref{conj:opt} is  based on the homogeneity of
sensors and symmetry of the problem setting (e.g., symmetric eigenvalues of 
$U$, uniform probability of final states, etc.). Thus, we believe the optimal
initial state solution for an unambiguous discrimination scheme is the same as in
the case of the minimum error scheme. Thus, Conjecture~\ref{conj:opt} should hold.

\item 
Conjecture~\ref{conj:avg} is independent of the measurement scheme.

\item 
We prove the version of Lemma~\ref{lemma:angle} corresponding to the unambiguous measurement below. Thus, Theorem~\ref{thm:final} also holds for unambiguous measurement. 

\item 
The optimization problem of determining the optimal measurement $\{\Pi_i\}$ for 
an unambiguous discrimination scheme can also be formulated as an SDP~\cite{Bae_2015}, 
and thus can be computed numerically. Thus, the search heuristics from \S\ref{sec:searching} will also work for unambiguous measurement with the 
corresponding SDP for an unambiguous discrimination scheme.

\end{enumerate}

\begin{lem-prf}
Consider $n$ states to be discriminated $\phi_0, \phi_1, \ldots, \phi_{n-1}$ 
such that $\bra{\phi_i}\ket{\phi_j} = x$,
for all $0 \leq i, j \leq n-1$ and $i \neq j$. 
The probability of {\bf failure} in discriminating 
$\phi_0, \phi_1, \ldots, \phi_{n-1}$ using an optimal measurement (for unambiguous
discrimination) 
increases with an increase in $x$ when $x \geq 0$.
\label{lemma:unambiguous}
\end{lem-prf}
\begin{prf}
The optimal/minimum probability of failure using the optimal POVM for a set of states with equal pairwise inner products is equal to $x$ when $x \geq 0$~\cite{quantum_pyramid}.
Thus, the lemma trivially holds. 
\end{prf}

\subsection{\bluee{Non-uniform Prior Probability}}
\bluee{Till now, we have implicitly assumed that the events (of affecting one sensor) occur with
a uniform probability. Here, we consider the generalization of allowing for the events to occur
with non-uniform probability. This could happen if different sensor locations can have different
probabilities of the event occurrence.} 

\bluee{
\para{Number of Sensors $n=2$}. 
When the number of sensors is 2, we observe that the optimal solution for the \iso problem actually remains unchanged. 
In particular, the expression for the minimum probability of error in discriminating the two final states, with
non-uniform probabilities $p_1$ and $p_2$, for a given initial state $|\psi\rangle$ is given by (derived from~\cite{helstrom}):  
\begin{equation}
P_{e} = \frac{1}{2} \left( 1 - \sqrt{ 1 - 4 p_1 p_2 |\langle \psi |(U\otimes U^{-1} ) |\psi\rangle |^{2}} \right). 
\label{eqn:two-prior}
\end{equation}
The above entails that, as for the case of uniform probabilities, we need to minimize $ |\langle \psi |(U\otimes U^{-1} ) |\psi\rangle |$, which is independent of $p_1$ and $p_2$. Thus, the optimal initial state for $n=2$ is independent of the probabilities associated with the final states/events.}

\begin{figure}[h]
    \centering
    \includegraphics[width=0.6\textwidth]{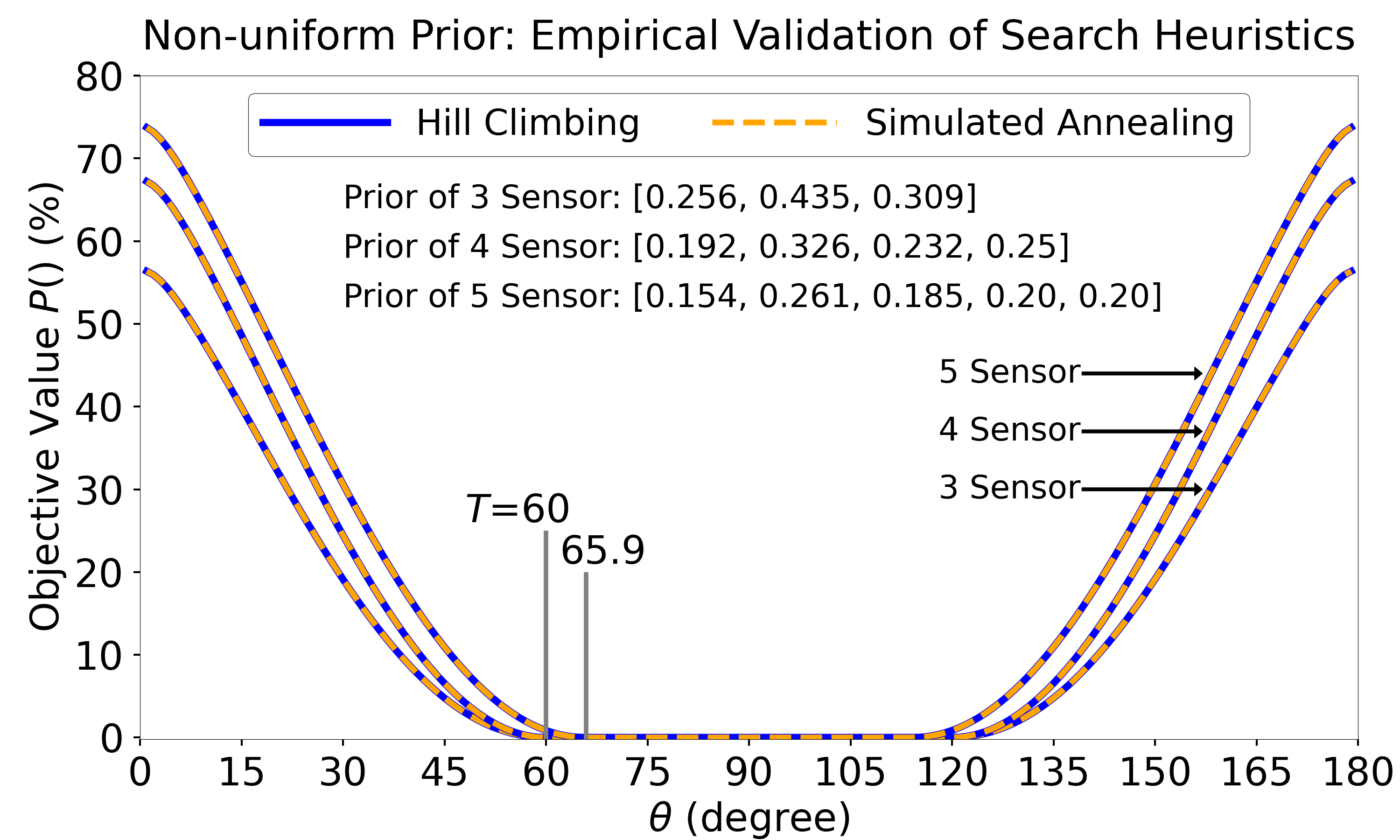}
    \caption{Performance of the three search heuristics with non-uniform prior for varying $U$'s parameter $\theta$, for a different number of sensors in the network. Genetic Algorithm (GA) is not shown explicitly, for clarity, but it also performs almost the same as Hill-Climbing and Simulated Annealing (SA), which are plotted above.}
    \label{fig:non-uniform-heuristics}
\end{figure}

\para{Number of Sensors $n > 2$.} 
\bluee{For $n>2$, it is easy to see that Theorems~\ref{thm:3sensor} and~\ref{thm:nsensors} that derive
conditions for orthogonality of the final states remain unchanged since the probabilities of 
events/final-states do not affect the final states themselves. 
However, the optimal \iso solution for general values of $\theta$ is certainly different
than that conjectured in Conjecture~\ref{conj:opt}, since Conjecture~\ref{conj:opt} is fundamentally
based on the symmetry of the final states, which is unlikely to be the case for non-uniform probabilities of
events.
On the other hand, it is easy to generalize the search heuristics for the case of non-uniform probability. See Fig.~\ref{fig:non-uniform-heuristics}, which plots the objective value $P()$ 
for varying $\theta$ for the three search heuristics. We observe that (i) The heuristics return
an optimal objective value (of zero) for the conditions in Theorems~\ref{thm:nsensors}; (ii) All the heuristics perform almost the same. These observations suggest that the heuristics likely perform near-optimally even for the 
general case of non-uniform event probabilities.
In addition, we note that, compared to the uniform probability case 
(i.e., Fig.~\ref{fig:heuristics}), the optimal objective value $P()$ 
under non-uniform probabilities is lower than the $P()$ under uniform probabilities, 
for any particular $\theta$.}

\subsection{\bluee{Impact of Quantum Noise.}}
\label{subsec:affect_noise}
\bluee{Till now, we have looked at the \iso problem from a theoretical perspective while ignoring the quantum noise. Since quantum noise is an essential aspect of quantum systems, we present a mitigation strategy to correct for quantum noise and evaluate it for two quantum noise models.}

\para{Quantum Noise-Mitigation Strategy.}
\bluee{In our context, the impact of the noise is that it essentially results in final states that are different (due to noise) than the ones we try to discriminate. 
That is, 
consider an given initial state $\ket{\psi}$ which yields (noiseless) final states $\{ \ket{\phi_i}\}$; let the optimal measurement to discriminate the final states $\{ \ket{\phi_i}\}$ be the POVM with elements $\{ E_i\}$.
However, due to the noise, the actual noisy final states may actually be different than $\{ \ket{\phi_i}\}$, which, when discriminated with the POVM $\{ E_i\}$, will result
in a higher probability of error than if there were no noise.
Thus, to account for such quantum noise, we propose to modify the POVM measurement appropriately. In particular, we compute the POVM measurement to discriminate 
the expected noisy final states---which we represent by the density matrices of the 
mixed states representing the ensemble of potential final states. 
More formally, our strategy is as follows: For each final state $\ket{\phi_i}$, let $\rho_i$ be the density matrix that represents the distribution/mixture of noisy final states that may
result instead of $\ket{\phi_i}$. 
Then, we use SDP (Eqn.~\ref{eqn:measure-sdp}) to determine the optimal POVM $\{ E'_i\}$ that optimally discriminates the density matrices $\{\rho_i\}$, 
and use it to discriminate the noisy final state.}

\begin{figure}[b]
    \centering
    \includegraphics[width=0.98\textwidth]{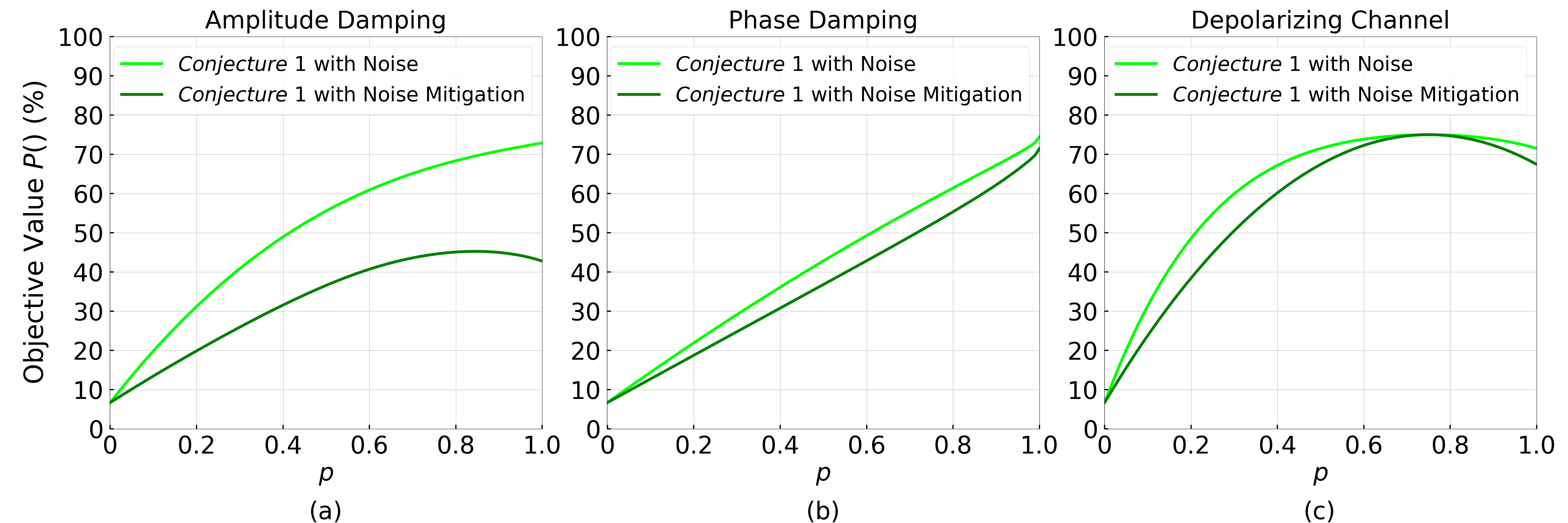}
    \caption{\bluee{The improvement in the objective value $P()$ for the Conjecture~1's solution due to the noise-mitigation strategy, for the three noise models, for $\theta=45$ degrees and 4 sensors.}}
    \label{fig:povm_noise}
\end{figure}

\bluee{
\para{Evaluation.}
We consider three popular noise models~\cite{nielsen2010quantum} for evaluation of our above mitigation technique. 
\begin{enumerate}
    \item \emph{Amplitude damping} causes the quantum system to lose energy.
    \item \emph{Phase damping} describes the loss of quantum information without energy loss.
    \item \emph{Depolarizing channel} is probabilistically replacing the qubit by the completely mixed state, $I/2$.
\end{enumerate}
All the above noise models can be characterized using the Kraus operators ($K$), which obey $\sum_{i} K_{i}^{\dagger}K_i=I$. In particular, the Kraus operators for the amplitude damping are:
$$ \mathcal{N}_{amp} = \{K_{a0}, K_{a1}\} = \{ \begin{bmatrix}
  1 & 0 \\
  0 & \sqrt{1 - p} \\
\end{bmatrix}, 
\begin{bmatrix}
  0 & \sqrt{p} \\
  0 & 0 \\
\end{bmatrix} \}$$
where $p$ can be thought of as the probability of losing a photon~\cite{nielsen2010quantum}. 
The Kraus operators for the phase damping are:
$$ \mathcal{N}_{pha} =   \{K_{p0}, K_{p1}\} = \{ \begin{bmatrix}
  1 & 0 \\
  0 & \sqrt{1 - p} \\
\end{bmatrix}, 
\begin{bmatrix}
  0 & 0 \\
  0 & \sqrt{p} \\
\end{bmatrix} \}$$
where $p$ can be interpreted as the probability that a photon from the system has been scattered (without loss of energy)~\cite{nielsen2010quantum}.
Finally, the Kraus operators for the depolarizing channel are:
$$ \mathcal{N}_{dep} = \{K_{d0}, K_{d1}, K_{d2}, K_{d3}\} =  \{\sqrt{1-p}\begin{bmatrix}
  1 & 0 \\
  0 & 1 \\
\end{bmatrix}, \sqrt{\frac{p}{3}}\begin{bmatrix}
  0 & 1 \\
  1 & 0 \\
\end{bmatrix}, \sqrt{\frac{p}{3}}\begin{bmatrix}
  0 & -i \\
  i & 0 \\
\end{bmatrix}, \sqrt{\frac{p}{3}}\begin{bmatrix}
  1 & 0 \\
  0 & -1 \\
\end{bmatrix} \}$$
where $p$ is the probability of a qubit being depolarized.
For a given noise model, its Kraus operators give the operators by which the state's density matrix is transformed with a corresponding probability. 
For example, in our context, under the third noise model of depolarizing noise, for a given initial state $\ket{\psi}$, each final state $\ket{\phi_i}$ with a density matrix $\rho_i = \ketbra{\phi_i}$ is transformed to $K_{d0}\rho_i K^{\dagger}_{d0}$ with a probability of $(1-p)$ and to $K_{d1}\rho_i K^{\dagger}_{d1}$ or $K_{d2}\rho_i K^{\dagger}_{d2}$ or 
$K_{d3}\rho_i K^{\dagger}_{d3}$ with a 
probability of $p/3$ each.
The above noise models are for a single sensor/qubit; 
for multiple qubits, we use a tensor product of the single-qubit noises.
Fig.~\ref{fig:povm_noise} shows the 1) impact of various quantum noise on the results, and 2) for the initial state (from Conjecture~1), how the objective value $P()$ improves due to the above-discussed noise-mitigation strategy for the three noise models, for the specific value of $\theta = 45$ degrees and the number of sensors equal to 4.
We observe that the improvement is particularly significant in the case of amplitude-damping noise.
}

\section{Conclusion and Future Directions}
\label{sec:conclusion}

In this work, we formulate the problem of initial state optimization in detector quantum sensor networks, which has potential applications in event localization.
We first derive the necessary and sufficient conditions for the existence of an initial state that can detect the firing sensor with perfect accuracy, i.e., with zero probability of error.
Using the insights from this result, we derive a conjectured optimal solution for the problem and provide a pathway to proving the conjecture.
Multiple search-based heuristics are also developed for the problem and the heuristics' numerical results successfully validate the conjecture.
In the end, we extend our results to the unambiguous discrimination measurement scheme, non-uniform prior, and considering quantum noise.

Beyond proving the stated Conjectures in the paper, there are many generalizations of the addressed
\iso problem of great interest in terms of: 
(i) More general final states (e.g, two sensors may change at a time, allowing for multiple impact operators $U_1, U_2$, etc.), 
(ii) Restricting the measurement operators allowed (e.g., allowing only the projective measurements and/or local measurements~\cite{umd-entanglement}), to incorporate practical considerations in the implementation of measurement operators. 
We are also interested in proving related results of interest, e.g., \iso initial-state solution being the same for minimum error and unambiguous discrimination.

\bibliographystyle{ACM-Reference-Format}
\bibliography{bib}


\begin{thebibliography}{39}


\ifx \showCODEN    \undefined \def \showCODEN     #1{\unskip}     \fi
\ifx \showDOI      \undefined \def \showDOI       #1{#1}\fi
\ifx \showISBNx    \undefined \def \showISBNx     #1{\unskip}     \fi
\ifx \showISBNxiii \undefined \def \showISBNxiii  #1{\unskip}     \fi
\ifx \showISSN     \undefined \def \showISSN      #1{\unskip}     \fi
\ifx \showLCCN     \undefined \def \showLCCN      #1{\unskip}     \fi
\ifx \shownote     \undefined \def \shownote      #1{#1}          \fi
\ifx \showarticletitle \undefined \def \showarticletitle #1{#1}   \fi
\ifx \showURL      \undefined \def \showURL       {\relax}        \fi
\providecommand\bibfield[2]{#2}
\providecommand\bibinfo[2]{#2}
\providecommand\natexlab[1]{#1}
\providecommand\showeprint[2][]{arXiv:#2}

\bibitem[Altenburg and Wölk(2018)]%
        {Altenburg_2019}
\bibfield{author}{\bibinfo{person}{Sanah Altenburg} {and} \bibinfo{person}{Sabine Wölk}.} \bibinfo{year}{2018}\natexlab{}.
\newblock \showarticletitle{Multi-parameter estimation: global, local and sequential strategies}.
\newblock \bibinfo{journal}{\emph{Physica Scripta}} \bibinfo{volume}{94}, \bibinfo{number}{1} (\bibinfo{date}{nov} \bibinfo{year}{2018}), \bibinfo{pages}{014001}.
\newblock
\urldef\tempurl%
\url{https://doi.org/10.1088/1402-4896/aaeca1}
\showDOI{\tempurl}


\bibitem[Assouly et~al\mbox{.}(2023)]%
        {quantum_radar}
\bibfield{author}{\bibinfo{person}{R Assouly}, \bibinfo{person}{R. Dassonneville}, \bibinfo{person}{T. Peronnin}, \bibinfo{person}{A. Bienfait}, {and} \bibinfo{person}{B. Huard}.} \bibinfo{year}{2023}\natexlab{}.
\newblock \showarticletitle{Quantum advantage in microwave quantum radar}.
\newblock \bibinfo{journal}{\emph{Nature Physics}} (\bibinfo{year}{2023}).
\newblock
\urldef\tempurl%
\url{https://doi.org/10.1038/s41567-023-02113-4}
\showDOI{\tempurl}


\bibitem[Bae and Kwek(2015)]%
        {Bae_2015}
\bibfield{author}{\bibinfo{person}{Joonwoo Bae} {and} \bibinfo{person}{Leong-Chuan Kwek}.} \bibinfo{year}{2015}\natexlab{}.
\newblock \showarticletitle{Quantum state discrimination and its applications}.
\newblock \bibinfo{journal}{\emph{Journal of Physics A: Mathematical and Theoretical}} \bibinfo{volume}{48}, \bibinfo{number}{8} (\bibinfo{date}{jan} \bibinfo{year}{2015}), \bibinfo{pages}{083001}.
\newblock
\urldef\tempurl%
\url{https://doi.org/10.1088/1751-8113/48/8/083001}
\showDOI{\tempurl}


\bibitem[Barnett and Croke(2009)]%
        {Barnett-review}
\bibfield{author}{\bibinfo{person}{Stephen~M. Barnett} {and} \bibinfo{person}{Sarah Croke}.} \bibinfo{year}{2009}\natexlab{}.
\newblock \showarticletitle{Quantum state discrimination}.
\newblock \bibinfo{journal}{\emph{Adv. Opt. Photon.}} \bibinfo{volume}{1}, \bibinfo{number}{2} (\bibinfo{date}{Apr} \bibinfo{year}{2009}), \bibinfo{pages}{238--278}.
\newblock
\urldef\tempurl%
\url{https://doi.org/10.1364/AOP.1.000238}
\showDOI{\tempurl}


\bibitem[B{\"a}rtschi and Eidenbenz(2019)]%
        {dicke_state}
\bibfield{author}{\bibinfo{person}{Andreas B{\"a}rtschi} {and} \bibinfo{person}{Stephan Eidenbenz}.} \bibinfo{year}{2019}\natexlab{}.
\newblock \showarticletitle{Deterministic Preparation of Dicke States}. In \bibinfo{booktitle}{\emph{Fundamentals of Computation Theory}}, \bibfield{editor}{\bibinfo{person}{Leszek~Antoni G{\k{a}}sieniec}, \bibinfo{person}{Jesper Jansson}, {and} \bibinfo{person}{Christos Levcopoulos}} (Eds.). \bibinfo{publisher}{Springer International Publishing}, \bibinfo{address}{Cham}, \bibinfo{pages}{126--139}.
\newblock
\showISBNx{978-3-030-25027-0}
\urldef\tempurl%
\url{https://doi.org/10.1007/978-3-030-25027-0_9}
\showURL{%
\tempurl}


\bibitem[Bergou(2007)]%
        {bergou-review-2007}
\bibfield{author}{\bibinfo{person}{János~A Bergou}.} \bibinfo{year}{2007}\natexlab{}.
\newblock \showarticletitle{Quantum state discrimination and selected applications}.
\newblock \bibinfo{journal}{\emph{Journal of Physics: Conference Series}} \bibinfo{volume}{84}, \bibinfo{number}{1} (\bibinfo{date}{oct} \bibinfo{year}{2007}), \bibinfo{pages}{012001}.
\newblock
\urldef\tempurl%
\url{https://doi.org/10.1088/1742-6596/84/1/012001}
\showDOI{\tempurl}


\bibitem[Bergou et~al\mbox{.}(2004)]%
        {bergou2004}
\bibfield{author}{\bibinfo{person}{J{\'a}nos~A. Bergou}, \bibinfo{person}{Ulrike Herzog}, {and} \bibinfo{person}{Mark Hillery}.} \bibinfo{year}{2004}\natexlab{}.
\newblock \bibinfo{booktitle}{\emph{11 Discrimination of Quantum States}}.
\newblock \bibinfo{publisher}{Springer Berlin Heidelberg}, \bibinfo{address}{Berlin, Heidelberg}, \bibinfo{pages}{417--465}.
\newblock
\showISBNx{978-3-540-44481-7}
\urldef\tempurl%
\url{https://doi.org/10.1007/978-3-540-44481-7_11}
\showDOI{\tempurl}


\bibitem[Bringewatt et~al\mbox{.}(2021)]%
        {PhysRevResearch.qsn}
\bibfield{author}{\bibinfo{person}{Jacob Bringewatt}, \bibinfo{person}{Igor Boettcher}, \bibinfo{person}{Pradeep Niroula}, \bibinfo{person}{Przemyslaw Bienias}, {and} \bibinfo{person}{Alexey~V. Gorshkov}.} \bibinfo{year}{2021}\natexlab{}.
\newblock \showarticletitle{Protocols for estimating multiple functions with quantum sensor networks: Geometry and performance}.
\newblock \bibinfo{journal}{\emph{Phys. Rev. Res.}}  \bibinfo{volume}{3} (\bibinfo{date}{Jul} \bibinfo{year}{2021}), \bibinfo{pages}{033011}.
\newblock
Issue 3.
\urldef\tempurl%
\url{https://doi.org/10.1103/PhysRevResearch.3.033011}
\showDOI{\tempurl}


\bibitem[Cao and Wu(2023)]%
        {umd-entanglement}
\bibfield{author}{\bibinfo{person}{Yingkang Cao} {and} \bibinfo{person}{Xiaodi Wu}.} \bibinfo{year}{2023}\natexlab{}.
\newblock \showarticletitle{Distributed Quantum Sensing Network with Geographically Constrained Measurement Strategies}. In \bibinfo{booktitle}{\emph{IEEE International Conference on Acoustics, Speech and Signal Processing (ICASSP)}}. \bibinfo{publisher}{IEEE}, \bibinfo{address}{Greece}, \bibinfo{pages}{1--5}.
\newblock
\urldef\tempurl%
\url{https://doi.org/10.1109/ICASSP49357.2023.10096723}
\showDOI{\tempurl}


\bibitem[Degen et~al\mbox{.}(2017)]%
        {RevModPhys.quantumsensing}
\bibfield{author}{\bibinfo{person}{C.~L. Degen}, \bibinfo{person}{F. Reinhard}, {and} \bibinfo{person}{P. Cappellaro}.} \bibinfo{year}{2017}\natexlab{}.
\newblock \showarticletitle{Quantum sensing}.
\newblock \bibinfo{journal}{\emph{Rev. Mod. Phys.}}  \bibinfo{volume}{89} (\bibinfo{year}{2017}), \bibinfo{pages}{035002}.
\newblock
Issue 3.
\urldef\tempurl%
\url{https://doi.org/10.1103/RevModPhys.89.035002}
\showDOI{\tempurl}


\bibitem[Diamond and Boyd(2016)]%
        {diamond2016cvxpy}
\bibfield{author}{\bibinfo{person}{Steven Diamond} {and} \bibinfo{person}{Stephen Boyd}.} \bibinfo{year}{2016}\natexlab{}.
\newblock \showarticletitle{{CVXPY}: {A} {P}ython-embedded modeling language for convex optimization}.
\newblock \bibinfo{journal}{\emph{Journal of Machine Learning Research}} \bibinfo{volume}{17}, \bibinfo{number}{83} (\bibinfo{year}{2016}), \bibinfo{pages}{1--5}.
\newblock


\bibitem[D’Ariano et~al\mbox{.}(2002)]%
        {DAriano_2002}
\bibfield{author}{\bibinfo{person}{G~Mauro D’Ariano}, \bibinfo{person}{Paoloplacido~Lo Presti}, {and} \bibinfo{person}{Matteo G~A Paris}.} \bibinfo{year}{2002}\natexlab{}.
\newblock \showarticletitle{Improved discrimination of unitary transformations by entangled probes}.
\newblock \bibinfo{journal}{\emph{Journal of Optics B: Quantum and Semiclassical Optics}} \bibinfo{volume}{4}, \bibinfo{number}{4} (\bibinfo{date}{jul} \bibinfo{year}{2002}), \bibinfo{pages}{S273}.
\newblock
\urldef\tempurl%
\url{https://doi.org/10.1088/1464-4266/4/4/304}
\showDOI{\tempurl}


\bibitem[Ehrenberg et~al\mbox{.}(2023)]%
        {ehrenberg}
\bibfield{author}{\bibinfo{person}{Adam Ehrenberg}, \bibinfo{person}{Jacob Bringewatt}, {and} \bibinfo{person}{Alexey~V. Gorshkov}.} \bibinfo{year}{2023}\natexlab{}.
\newblock \showarticletitle{Minimum-entanglement protocols for function estimation}.
\newblock \bibinfo{journal}{\emph{Phys. Rev. Res.}}  \bibinfo{volume}{5} (\bibinfo{date}{Sep} \bibinfo{year}{2023}), \bibinfo{pages}{033228}.
\newblock
Issue 3.
\urldef\tempurl%
\url{https://doi.org/10.1103/PhysRevResearch.5.033228}
\showDOI{\tempurl}


\bibitem[Eldar et~al\mbox{.}(2003)]%
        {Eldar_2003}
\bibfield{author}{\bibinfo{person}{Y.C. Eldar}, \bibinfo{person}{A. Megretski}, {and} \bibinfo{person}{G.C. Verghese}.} \bibinfo{year}{2003}\natexlab{}.
\newblock \showarticletitle{Designing optimal quantum detectors via semidefinite programming}.
\newblock \bibinfo{journal}{\emph{{IEEE} Transactions on Information Theory}} \bibinfo{volume}{49}, \bibinfo{number}{4} (\bibinfo{date}{apr} \bibinfo{year}{2003}), \bibinfo{pages}{1007--1012}.
\newblock
\urldef\tempurl%
\url{https://doi.org/10.1109/tit.2003.809510}
\showDOI{\tempurl}


\bibitem[Eldredge et~al\mbox{.}(2018)]%
        {Eldredge_2018}
\bibfield{author}{\bibinfo{person}{Zachary Eldredge}, \bibinfo{person}{Michael Foss-Feig}, \bibinfo{person}{Jonathan~A. Gross}, \bibinfo{person}{S.~L. Rolston}, {and} \bibinfo{person}{Alexey~V. Gorshkov}.} \bibinfo{year}{2018}\natexlab{}.
\newblock \showarticletitle{Optimal and secure measurement protocols for quantum sensor networks}.
\newblock \bibinfo{journal}{\emph{Phys. Rev. A}}  \bibinfo{volume}{97} (\bibinfo{date}{Apr} \bibinfo{year}{2018}), \bibinfo{pages}{042337}.
\newblock
Issue 4.
\urldef\tempurl%
\url{https://doi.org/10.1103/PhysRevA.97.042337}
\showDOI{\tempurl}


\bibitem[Englert and {\v{R}}eh{\'a}{\v{c}}ek(2010)]%
        {quantum_pyramid}
\bibfield{author}{\bibinfo{person}{Berthold-Georg Englert} {and} \bibinfo{person}{Jaroslav {\v{R}}eh{\'a}{\v{c}}ek}.} \bibinfo{year}{2010}\natexlab{}.
\newblock \showarticletitle{How well can you know the edge of a quantum pyramid?}
\newblock \bibinfo{journal}{\emph{Journal of Modern Optics}} \bibinfo{volume}{57}, \bibinfo{number}{3} (\bibinfo{year}{2010}), \bibinfo{pages}{218--226}.
\newblock
\urldef\tempurl%
\url{https://doi.org/10.1080/09500340903094601}
\showURL{%
\tempurl}


\bibitem[Ge et~al\mbox{.}(2018)]%
        {Jacobs_2018}
\bibfield{author}{\bibinfo{person}{Wenchao Ge}, \bibinfo{person}{Kurt Jacobs}, \bibinfo{person}{Zachary Eldredge}, \bibinfo{person}{Alexey~V. Gorshkov}, {and} \bibinfo{person}{Michael Foss-Feig}.} \bibinfo{year}{2018}\natexlab{}.
\newblock \showarticletitle{Distributed Quantum Metrology with Linear Networks and Separable Inputs}.
\newblock \bibinfo{journal}{\emph{Phys. Rev. Lett.}}  \bibinfo{volume}{121} (\bibinfo{date}{Jul} \bibinfo{year}{2018}), \bibinfo{pages}{043604}.
\newblock
Issue 4.
\urldef\tempurl%
\url{https://doi.org/10.1103/PhysRevLett.121.043604}
\showDOI{\tempurl}


\bibitem[Giovannetti et~al\mbox{.}(2011)]%
        {Giovannetti_2011}
\bibfield{author}{\bibinfo{person}{Vittorio Giovannetti}, \bibinfo{person}{Seth Lloyd}, {and} \bibinfo{person}{Lorenzo Maccone}.} \bibinfo{year}{2011}\natexlab{}.
\newblock \showarticletitle{Advances in quantum metrology}.
\newblock \bibinfo{journal}{\emph{Nature Photonics}} \bibinfo{volume}{5}, \bibinfo{number}{4} (\bibinfo{date}{mar} \bibinfo{year}{2011}), \bibinfo{pages}{222--229}.
\newblock
\urldef\tempurl%
\url{https://doi.org/10.1038/nphoton.2011.35}
\showDOI{\tempurl}


\bibitem[Helstrom(1976)]%
        {helstrom}
\bibfield{author}{\bibinfo{person}{C.W. Helstrom}.} \bibinfo{year}{1976}\natexlab{}.
\newblock \bibinfo{booktitle}{\emph{Quantum Detection and Estimation Theory}}.
\newblock \bibinfo{publisher}{Academic Press}, \bibinfo{address}{New York}.
\newblock


\bibitem[Hillery et~al\mbox{.}(2023)]%
        {Hillery_2023}
\bibfield{author}{\bibinfo{person}{Mark Hillery}, \bibinfo{person}{Himanshu Gupta}, {and} \bibinfo{person}{Caitao Zhan}.} \bibinfo{year}{2023}\natexlab{}.
\newblock \showarticletitle{Discrete outcome quantum sensor networks}.
\newblock \bibinfo{journal}{\emph{Phys. Rev. A}}  \bibinfo{volume}{107} (\bibinfo{date}{Jan} \bibinfo{year}{2023}), \bibinfo{pages}{012435}.
\newblock
Issue 1.
\urldef\tempurl%
\url{https://doi.org/10.1103/PhysRevA.107.012435}
\showDOI{\tempurl}


\bibitem[Holland(1992)]%
        {Holland_1992}
\bibfield{author}{\bibinfo{person}{John~H. Holland}.} \bibinfo{year}{1992}\natexlab{}.
\newblock \showarticletitle{Genetic Algorithms}.
\newblock \bibinfo{journal}{\emph{Scientific American}} \bibinfo{volume}{267}, \bibinfo{number}{1} (\bibinfo{year}{1992}), \bibinfo{pages}{66--73}.
\newblock
\urldef\tempurl%
\url{http://www.jstor.org/stable/24939139}
\showURL{%
\tempurl}


\bibitem[Katoch et~al\mbox{.}(2021)]%
        {review-ga}
\bibfield{author}{\bibinfo{person}{Sourabh Katoch}, \bibinfo{person}{Sumit~Singh Chauhan}, {and} \bibinfo{person}{Vijay Kumar}.} \bibinfo{year}{2021}\natexlab{}.
\newblock \showarticletitle{A review on genetic algorithm: past, present, and future}.
\newblock \bibinfo{journal}{\emph{Multimedia Tools and Applications}}  \bibinfo{volume}{80} (\bibinfo{year}{2021}), \bibinfo{pages}{8091--8126}.
\newblock
\urldef\tempurl%
\url{https://doi.org/10.1007/s11042-020-10139-6}
\showURL{%
\tempurl}


\bibitem[Kirkpatrick et~al\mbox{.}(1983)]%
        {Kirkpatrick_1983}
\bibfield{author}{\bibinfo{person}{S. Kirkpatrick}, \bibinfo{person}{C.~D. Gelatt}, {and} \bibinfo{person}{M.~P. Vecchi}.} \bibinfo{year}{1983}\natexlab{}.
\newblock \showarticletitle{Optimization by Simulated Annealing}.
\newblock \bibinfo{journal}{\emph{Science}} \bibinfo{volume}{220}, \bibinfo{number}{4598} (\bibinfo{year}{1983}), \bibinfo{pages}{671--680}.
\newblock
\urldef\tempurl%
\url{https://doi.org/10.1126/science.220.4598.671}
\showDOI{\tempurl}


\bibitem[Koczor et~al\mbox{.}(2020)]%
        {Koczor_2020}
\bibfield{author}{\bibinfo{person}{Bálint Koczor}, \bibinfo{person}{Suguru Endo}, \bibinfo{person}{Tyson Jones}, \bibinfo{person}{Yuichiro Matsuzaki}, {and} \bibinfo{person}{Simon~C Benjamin}.} \bibinfo{year}{2020}\natexlab{}.
\newblock \showarticletitle{Variational-state quantum metrology}.
\newblock \bibinfo{journal}{\emph{New Journal of Physics}} \bibinfo{volume}{22}, \bibinfo{number}{8} (\bibinfo{date}{aug} \bibinfo{year}{2020}), \bibinfo{pages}{083038}.
\newblock
\urldef\tempurl%
\url{https://doi.org/10.1088/1367-2630/ab965e}
\showDOI{\tempurl}


\bibitem[Laarhoven and Aarts(1987)]%
        {Laarhoven_1987}
\bibfield{author}{\bibinfo{person}{P.~J.~M. Laarhoven} {and} \bibinfo{person}{E.~H.~L. Aarts}.} \bibinfo{year}{1987}\natexlab{}.
\newblock \bibinfo{booktitle}{\emph{Simulated Annealing: Theory and Applications}}.
\newblock \bibinfo{publisher}{Kluwer Academic Publishers}, \bibinfo{address}{USA}.
\newblock
\showISBNx{9027725136}


\bibitem[Lewis(2007)]%
        {SA-sudoku}
\bibfield{author}{\bibinfo{person}{Rhyd Lewis}.} \bibinfo{year}{2007}\natexlab{}.
\newblock \showarticletitle{Metaheuristics Can Solve Sudoku Puzzles}.
\newblock \bibinfo{journal}{\emph{Journal of Heuristics}}  \bibinfo{volume}{13} (\bibinfo{year}{2007}), \bibinfo{pages}{387--401}.
\newblock
\urldef\tempurl%
\url{https://doi.org/10.1007/s10732-007-9012-8}
\showURL{%
\tempurl}


\bibitem[Li et~al\mbox{.}(2020)]%
        {mpe_2020}
\bibfield{author}{\bibinfo{person}{Xinwei Li}, \bibinfo{person}{Jia-Hao Cao}, \bibinfo{person}{Qi Liu}, \bibinfo{person}{Meng~Khoon Tey}, {and} \bibinfo{person}{Li You}.} \bibinfo{year}{2020}\natexlab{}.
\newblock \showarticletitle{Multi-parameter estimation with multi-mode Ramsey interferometry}.
\newblock \bibinfo{journal}{\emph{New Journal of Physics}} \bibinfo{volume}{22}, \bibinfo{number}{4} (\bibinfo{date}{apr} \bibinfo{year}{2020}), \bibinfo{pages}{043005}.
\newblock
\urldef\tempurl%
\url{https://doi.org/10.1088/1367-2630/ab7a32}
\showDOI{\tempurl}


\bibitem[Nielsen and Chuang(2010)]%
        {nielsen2010quantum}
\bibfield{author}{\bibinfo{person}{Michael~A Nielsen} {and} \bibinfo{person}{Isaac~L Chuang}.} \bibinfo{year}{2010}\natexlab{}.
\newblock \bibinfo{booktitle}{\emph{Quantum computation and quantum information}}.
\newblock \bibinfo{publisher}{Cambridge university press}, \bibinfo{address}{Cambridge, UK}.
\newblock


\bibitem[O'Donoghue et~al\mbox{.}(2016)]%
        {sdp-solver}
\bibfield{author}{\bibinfo{person}{Brendan O'Donoghue}, \bibinfo{person}{Eric Chu}, \bibinfo{person}{Neal Parikh}, {and} \bibinfo{person}{Stephen Boyd}.} \bibinfo{year}{2016}\natexlab{}.
\newblock \showarticletitle{Conic Optimization via Operator Splitting and Homogeneous Self-Dual Embedding}.
\newblock \bibinfo{journal}{\emph{Journal of Optimization Theory and Applications}} \bibinfo{volume}{169}, \bibinfo{number}{3} (\bibinfo{date}{June} \bibinfo{year}{2016}), \bibinfo{pages}{1042--1068}.
\newblock
\urldef\tempurl%
\url{http://stanford.edu/~boyd/papers/scs.html}
\showURL{%
\tempurl}


\bibitem[Pirandola et~al\mbox{.}(2018)]%
        {photonic_quantum_sensing}
\bibfield{author}{\bibinfo{person}{S. Pirandola}, \bibinfo{person}{B.~R. Bardhan}, \bibinfo{person}{T. Gehring}, \bibinfo{person}{C. Weedbrook}, {and} \bibinfo{person}{S. Lloyd}.} \bibinfo{year}{2018}\natexlab{}.
\newblock \showarticletitle{Advances in photonic quantum sensing}.
\newblock \bibinfo{journal}{\emph{Nature Photonics}} (\bibinfo{year}{2018}).
\newblock
\urldef\tempurl%
\url{https://doi.org/10.1038/s41566-018-0301-6}
\showDOI{\tempurl}


\bibitem[Proctor et~al\mbox{.}(2018)]%
        {mpe_2018}
\bibfield{author}{\bibinfo{person}{Timothy~J. Proctor}, \bibinfo{person}{Paul~A. Knott}, {and} \bibinfo{person}{Jacob~A. Dunningham}.} \bibinfo{year}{2018}\natexlab{}.
\newblock \showarticletitle{Multiparameter Estimation in Networked Quantum Sensors}.
\newblock \bibinfo{journal}{\emph{Phys. Rev. Lett.}}  \bibinfo{volume}{120} (\bibinfo{date}{Feb} \bibinfo{year}{2018}), \bibinfo{pages}{080501}.
\newblock
Issue 8.
\urldef\tempurl%
\url{https://doi.org/10.1103/PhysRevLett.120.080501}
\showDOI{\tempurl}


\bibitem[Qian et~al\mbox{.}(2021)]%
        {PhysRevA.qsn}
\bibfield{author}{\bibinfo{person}{Timothy Qian}, \bibinfo{person}{Jacob Bringewatt}, \bibinfo{person}{Igor Boettcher}, \bibinfo{person}{Przemyslaw Bienias}, {and} \bibinfo{person}{Alexey~V. Gorshkov}.} \bibinfo{year}{2021}\natexlab{}.
\newblock \showarticletitle{Optimal measurement of field properties with quantum sensor networks}.
\newblock \bibinfo{journal}{\emph{Phys. Rev. A}}  \bibinfo{volume}{103} (\bibinfo{date}{Mar} \bibinfo{year}{2021}), \bibinfo{pages}{L030601}.
\newblock
Issue 3.
\urldef\tempurl%
\url{https://doi.org/10.1103/PhysRevA.103.L030601}
\showDOI{\tempurl}


\bibitem[Roga et~al\mbox{.}(2023)]%
        {dickestate_distributed}
\bibfield{author}{\bibinfo{person}{Wojciech Roga}, \bibinfo{person}{Rikizo Ikuta}, \bibinfo{person}{Tomoyuki Horikiri}, {and} \bibinfo{person}{Masahiro Takeoka}.} \bibinfo{year}{2023}\natexlab{}.
\newblock \showarticletitle{Efficient Dicke-state distribution in a network of lossy channels}.
\newblock \bibinfo{journal}{\emph{Phys. Rev. A}}  \bibinfo{volume}{108} (\bibinfo{date}{Jul} \bibinfo{year}{2023}), \bibinfo{pages}{012612}.
\newblock
Issue 1.
\urldef\tempurl%
\url{https://doi.org/10.1103/PhysRevA.108.012612}
\showDOI{\tempurl}


\bibitem[Rubio et~al\mbox{.}(2020)]%
        {Rubio_2020}
\bibfield{author}{\bibinfo{person}{Jesús Rubio}, \bibinfo{person}{Paul~A Knott}, \bibinfo{person}{Timothy~J Proctor}, {and} \bibinfo{person}{Jacob~A Dunningham}.} \bibinfo{year}{2020}\natexlab{}.
\newblock \showarticletitle{Quantum sensing networks for the estimation of linear functions}.
\newblock \bibinfo{journal}{\emph{Journal of Physics A: Mathematical and Theoretical}} \bibinfo{volume}{53}, \bibinfo{number}{34} (\bibinfo{date}{aug} \bibinfo{year}{2020}), \bibinfo{pages}{344001}.
\newblock
\urldef\tempurl%
\url{https://doi.org/10.1088/1751-8121/ab9d46}
\showDOI{\tempurl}


\bibitem[Saleem et~al\mbox{.}(2023)]%
        {saleem_dickestate}
\bibfield{author}{\bibinfo{person}{Zain~H. Saleem}, \bibinfo{person}{Michael Perlin}, \bibinfo{person}{Anil Shaji}, {and} \bibinfo{person}{Stephen~K. Gray}.} \bibinfo{year}{2023}\natexlab{}.
\newblock \bibinfo{title}{Achieving the Heisenberg limit with Dicke states in noisy quantum metrology}.
\newblock
\newblock
\showeprint[arxiv]{2309.12411}~[quant-ph]


\bibitem[Sidhu and Kok(2020)]%
        {Sidhu_2020}
\bibfield{author}{\bibinfo{person}{Jasminder~S. Sidhu} {and} \bibinfo{person}{Pieter Kok}.} \bibinfo{year}{2020}\natexlab{}.
\newblock \showarticletitle{{Geometric perspective on quantum parameter estimation}}.
\newblock \bibinfo{journal}{\emph{AVS Quantum Science}} \bibinfo{volume}{2}, \bibinfo{number}{1} (\bibinfo{date}{02} \bibinfo{year}{2020}), \bibinfo{pages}{014701}.
\newblock
\showISSN{2639-0213}
\urldef\tempurl%
\url{https://doi.org/10.1116/1.5119961}
\showDOI{\tempurl}


\bibitem[Suzuki et~al\mbox{.}(2020)]%
        {Suzuki_2020}
\bibfield{author}{\bibinfo{person}{Jun Suzuki}, \bibinfo{person}{Yuxiang Yang}, {and} \bibinfo{person}{Masahito Hayashi}.} \bibinfo{year}{2020}\natexlab{}.
\newblock \showarticletitle{Quantum state estimation with nuisance parameters}.
\newblock \bibinfo{journal}{\emph{Journal of Physics A: Mathematical and Theoretical}} \bibinfo{volume}{53}, \bibinfo{number}{45} (\bibinfo{date}{oct} \bibinfo{year}{2020}), \bibinfo{pages}{453001}.
\newblock
\urldef\tempurl%
\url{https://doi.org/10.1088/1751-8121/ab8b78}
\showDOI{\tempurl}


\bibitem[Zhan(2023)]%
        {iso-code}
\bibfield{author}{\bibinfo{person}{Caitao Zhan}.} \bibinfo{year}{2023}\natexlab{}.
\newblock \bibinfo{title}{Source Code Repository}.
\newblock
\newblock
\newblock
\shownote{\url{https://github.com/caitaozhan/QuantumSensorNetwork}}.


\bibitem[Zhuang and Zhang(2019)]%
        {slaen}
\bibfield{author}{\bibinfo{person}{Quntao Zhuang} {and} \bibinfo{person}{Zheshen Zhang}.} \bibinfo{year}{2019}\natexlab{}.
\newblock \showarticletitle{Physical-layer supervised learning assisted by an entangled sensor network}.
\newblock \bibinfo{journal}{\emph{Physical Review X}} \bibinfo{volume}{9}, \bibinfo{number}{4} (\bibinfo{year}{2019}), \bibinfo{pages}{041023}.
\newblock
\urldef\tempurl%
\url{https://link.aps.org/doi/10.1103/PhysRevX.9.041023}
\showURL{%
\tempurl}


\end{thebibliography}
\end{document}